\documentclass[a4paper,11pt]{article}
\pdfoutput=1 
\usepackage{jheppub} 

\usepackage[T1]{fontenc} 

\RequirePackage[displaymath]{lineno} 
\usepackage{epstopdf}
\usepackage{epsfig}
\usepackage{graphicx}
\usepackage{dcolumn}
\usepackage{bm}
\usepackage{ltablex,booktabs}
\usepackage{overpic}
\usepackage{subfigure}
\usepackage{float}
\usepackage{color}
\usepackage{amsmath}
\usepackage{mathcomp}
\usepackage{mathrsfs}
\usepackage{multirow}
\usepackage{rotating}
\usepackage{amssymb}
\usepackage{gensymb}
\usepackage{amsmath}
\usepackage{tabularx}
\usepackage{threeparttable}

\title{Observation of the decays $\chi_{cJ}\to n K^{0}_{S}\bar\Lambda + c.c.$}

\collaborationImg{\includegraphics[width=0.15\textwidth, angle=90]{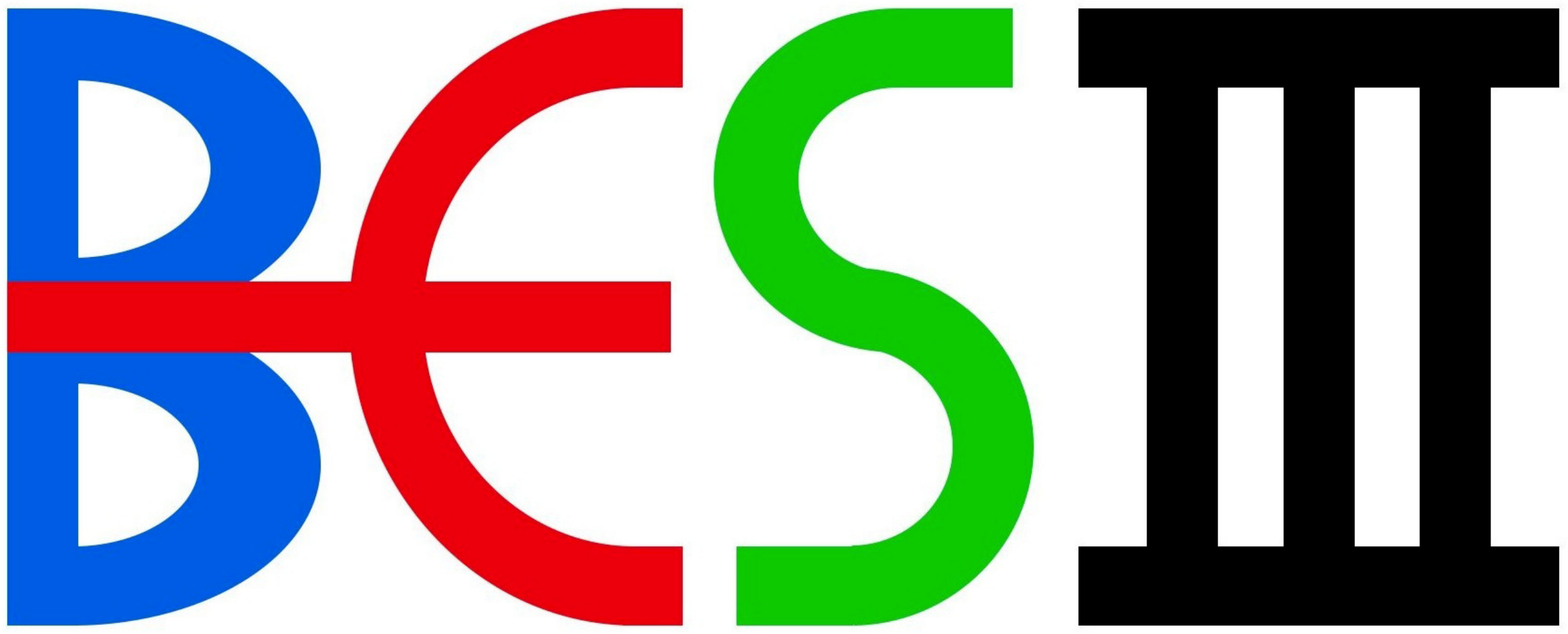}}
\collaboration{BESIII Collaboration}

\author{
M.~Ablikim$^{1}$, M.~N.~Achasov$^{10,b}$, P.~Adlarson$^{67}$, S. ~Ahmed$^{15}$, M.~Albrecht$^{4}$, R.~Aliberti$^{28}$, A.~Amoroso$^{66A,66C}$, M.~R.~An$^{32}$, Q.~An$^{63,49}$, X.~H.~Bai$^{57}$
% ...
\\
\vspace{0.2cm} {\it
$^{1}$ Institute of High Energy Physics, Beijing 100049, People's Republic of China\\
% ...
$^{4}$ Bochum Ruhr-University, D-44780 Bochum, Germany\\
% ...
$^{10}$ G.I. Budker Institute of Nuclear Physics SB RAS (BINP), Novosibirsk 630090, Russia\\
% ...
$^{15}$ Helmholtz Institute Mainz, Staudinger Weg 18, D-55099 Mainz, Germany\\
% ...
$^{28}$ Johannes Gutenberg University of Mainz, Johann-Joachim-Becher-Weg 45, D-55099 Mainz, Germany\\
% ...
$^{32}$ Liaoning Normal University, Dalian 116029, People's Republic of China\\
% ...
$^{49}$ State Key Laboratory of Particle Detection and Electronics, Beijing 100049, Hefei 230026, People's Republic of China\\
% ...
$^{57}$ University of Jinan, Jinan 250022, People's Republic of China\\
% ...
$^{63}$ University of Science and Technology of China, Hefei 230026, People's Republic of China\\
% ...
$^{66}$ University of Turin and INFN, (A)University of Turin, I-10125, Turin, Italy; (B)University of Eastern Piedmont, I-15121, Alessandria, Italy; (C)INFN, I-10125, Turin, Italy\\
$^{67}$ Uppsala University, Box 516, SE-75120 Uppsala, Sweden\\
% ...
$^{b}$ Also at the Novosibirsk State University, Novosibirsk, 630090, Russia\\
% ...
}

}

\abstract{ By analyzing $4.48\times10^8$ $\psi(3686)$ events collected
  with the BESIII detector, we observe the decays $\chi_{cJ} \to n
  K^0_S\bar\Lambda + c.c.$~($J=0$,~1,~2) for the first time, via the
  radiative transition $\psi(3686) \to \gamma \chi_{cJ}$.  The
  branching fractions are determined to be $(6.67 \pm 0.26_{\rm stat}
  \pm 0.41_{\rm syst})\times10^{-4}$, $(1.71 \pm 0.12_{\rm stat} \pm
  0.12_{\rm syst})\times10^{-4}$, and $(3.66 \pm 0.17_{\rm stat} \pm
  0.23_{\rm syst})\times10^{-4}$ for $J=0$, 1, and 2, respectively.
}
\arxivnumber{}

\begin{document}
\maketitle
\flushbottom

\section{Introduction}
Studying the hadronic decays of the $c\bar{c}$ states $J/\psi$,
$\psi(3686)$, and $\chi_{cJ}$ ($J=0$,~1,~2) provides valuable
information on perturbative QCD in the charmonium-mass regime and on
the structure of charmonia.  The color-octet mechanism, which
successfully describes several decay patterns of P-wave $\chi_{cJ}$
states~\cite{ref::Wong}, may be applicable to further $\chi_{cJ}$
decays.  Measurements of $\chi_{cJ}$ hadronic decays can provide new
input on the color-octet mechanism and further assist in understanding
$\chi_{cJ}$ decay mechanisms.

The BES Collaboration observed near-threshold structures in
baryon-antibaryon invariant mass distributions in the radiative decay
$J/\psi\to\gamma p\bar{p}$~\cite{ref::prl-91-022001} and the purely
hadronic decay $J/\psi\to
pK^-\bar{\Lambda}$~\cite{ref::prl-93-112002}.  It was
suggested theoretically that these near-threshold structures might be
observation of baryonium~\cite{ref::Datta1, ref::Datta2, ref::Datta3}
or caused by final state interactions~\cite{ref::Kerbikov1, ref::Kerbikov2, ref::Kerbikov3}.
 Excited $\Lambda$ and
$N$ resonances were observed in the study of the decay
$J/\psi\to nK_S^0\bar{\Lambda}$~\cite{ref::plb-659-789}.
 Studying the same decay modes in
other charmonia can provide complementary information on these
structures.

Anomalous enhancements near the threshold of $p\bar{\Lambda}+c.c.$
have been also observed in the decays of $\chi_{cJ}\to p
K^-\bar\Lambda +c.c.$ by the BESIII
Collaboration~\cite{ref::paper-pkl}. If isospin symmetry is
conserved, the decay branching fraction~(BF) ratio
$\mathcal{B}(\chi_{cJ}\to p K^-\bar\Lambda+c.c.)/\mathcal{B}(\chi_{cJ}\to n K^0_S\bar\Lambda+c.c.)\sim 2$ should
be satisfied, thereby implying the existence of the isospin conjugate
decays $\chi_{cJ}\to nK^0_S\bar \Lambda+c.c.$.

In this analysis, we present the study of $\chi_{cJ}\to n
K^0_S\bar\Lambda+c.c.$ using the $\psi(3686)$ data sample containing
$(4.48\pm 0.03)\times 10^{8}$ events collected at
BESIII~\cite{ref::CPC42_023001}.  The radiative decays
$\psi(3686) \to \gamma\chi_{cJ}$, which have a BF
of approximately 10\% for each $\chi_{cJ}$~\cite{ref::pdg2014}, offer
an ideal environment to investigate $\chi_{cJ}$ decays. Throughout
this paper, charge conjugate modes are implied unless otherwise
stated.

\section{Detector and data sets}
\label{sec:detector_dataset}
The BESIII detector is a magnetic spectrometer~\cite{Ablikim:2009aa, Ablikim:2019hff}
located at the Beijing Electron Positron Collider~(BEPCII)~\cite{Yu:IPAC2016-TUYA01}.
The cylindrical core of the BESIII detector consists of a helium-based
multilayer drift chamber~(MDC), a plastic scintillator time-of-flight
system~(TOF), and a CsI(Tl) electromagnetic calorimeter~(EMC), which are all
enclosed in a superconducting solenoidal magnet providing a 1.0~T magnetic
field. The solenoid is supported by an octagonal flux-return yoke with
resistive plate counter muon identifier modules interleaved with steel. The
acceptance of charged particles and photons is 93\% over $4\pi$ solid angle.
The charged-particle momentum resolution at 1.0~GeV/$c$ is $0.5\%$, and the
specific energy loss~($dE/dx$) resolution is $6\%$ for the electrons from
Bhabha scattering. The EMC measures photon energies with a resolution of
$2.5\%$~($5\%$) at $1$~GeV in the barrel (end cap) region. The time resolution
of the TOF barrel part is 68~ps, while that of the end cap part is 110~ps.

Simulated samples produced with the {\sc geant4}-based~\cite{GEANT4}
Monte-Carlo~(MC) software, which includes the geometric description of
the BESIII detector and the detector response, are used to determine
the detection efficiencies and to estimate the background levels. The
simulation takes into account the beam energy spread and initial state
radiation~(ISR) in the $e^+e^-$ annihilation modeled with the
generator {\sc kkmc}~\cite{KKMC}. The inclusive MC samples consist of
$5.06\times10^{8}$ $\psi(3686)$ events, the ISR production of the
$J/\psi$ state, and the continuum processes incorporated in {\sc
  kkmc}. The known decay modes are modeled with {\sc
  eventgen}~\cite{EVTGEN1} using the BFs taken from the
Particle Data Group~(PDG)~\cite{ref::pdg2014}, and the remaining
unknown decays from the charmonium states with {\sc
  lundcharm}~\cite{LUNDCHARM1}.  Radiation from charged
final state particles is incorporated with the {\sc
  photos}~\cite{PHOTOS} package.

The signal detection efficiencies are estimated with signal MC
samples. The decays of $\psi(3686)\to \gamma\chi_{cJ}$ ($J=0$,~1,~2)
are simulated following Ref.~\cite{ref::paper-ggJp}, in which the
magnetic-quadrupole~(M2) transition for $\psi(3686)\to
\gamma\chi_{c1,2}$ and the electricoctupole~(E3) transition for
$\psi(3686)\to \gamma\chi_{c2}$ are considered in addition to the
dominant electric-dipole~(E1) transition. For the decay $\chi_{cJ}\to
nK^0_S\bar\Lambda+c.c.$, a special generator based on results of
Helicity Partial Wave Analysis
(HelPWA)~\cite{SM, ref::chung,ref::blatt,ref::helb3} is used.  The
background MC samples are obtained from inclusive MC samples, in which
the signal channels are removed with TopoAna~\cite{TopoAna}, and the
samples are normalized to the data sample based on luminosity.

\section{Event selection and background analysis}
\label{chap:event_selection}
The signal process $\psi(3686)\to \gamma\chi_{cJ}, \chi_{cJ}\to n
K^0_S\bar\Lambda$ with $K^{0}_{S}\to\pi^+\pi^-$ and $\bar\Lambda \to
\bar{p}\pi^{+}$ consists of the final state particles $\gamma
n\bar{p}\pi^{+}\pi^{+}\pi^{-}$.  Charged track candidates from the MDC
must satisfy $\vert\!\cos\theta\vert<0.93$, where $\theta$ is the
polar angle with respect to the $z$ axis, which is the axis of the
MDC. The closest approach to the interaction point is required to be
less than 20~cm along the $z$ direction and less than 10~cm in the
plane perpendicular to $z$. The TOF and $dE/dx$ information are
combined to calculate the particle identification probabilities ($P$)
for the hypotheses that a track is a pion, kaon, or proton. Proton
candidates are required to satisfy $P(p)>P(K)$ and
$P(p)>P(\pi)$. Exactly four charged tracks are required in each
candidate event.

\begin{figure}[htbp]
\begin{center}
\includegraphics[width=0.32\textwidth]{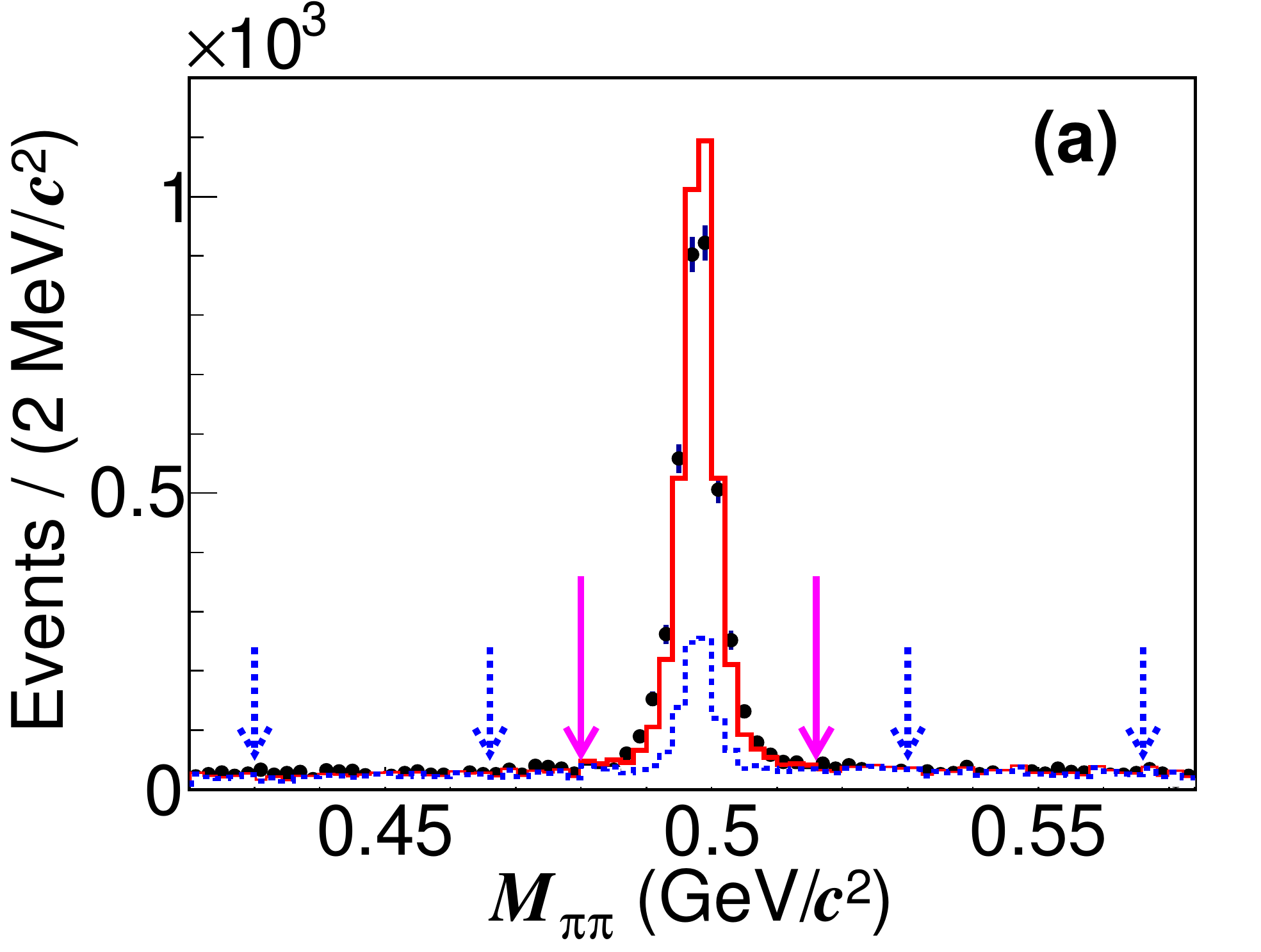}
\includegraphics[width=0.32\textwidth]{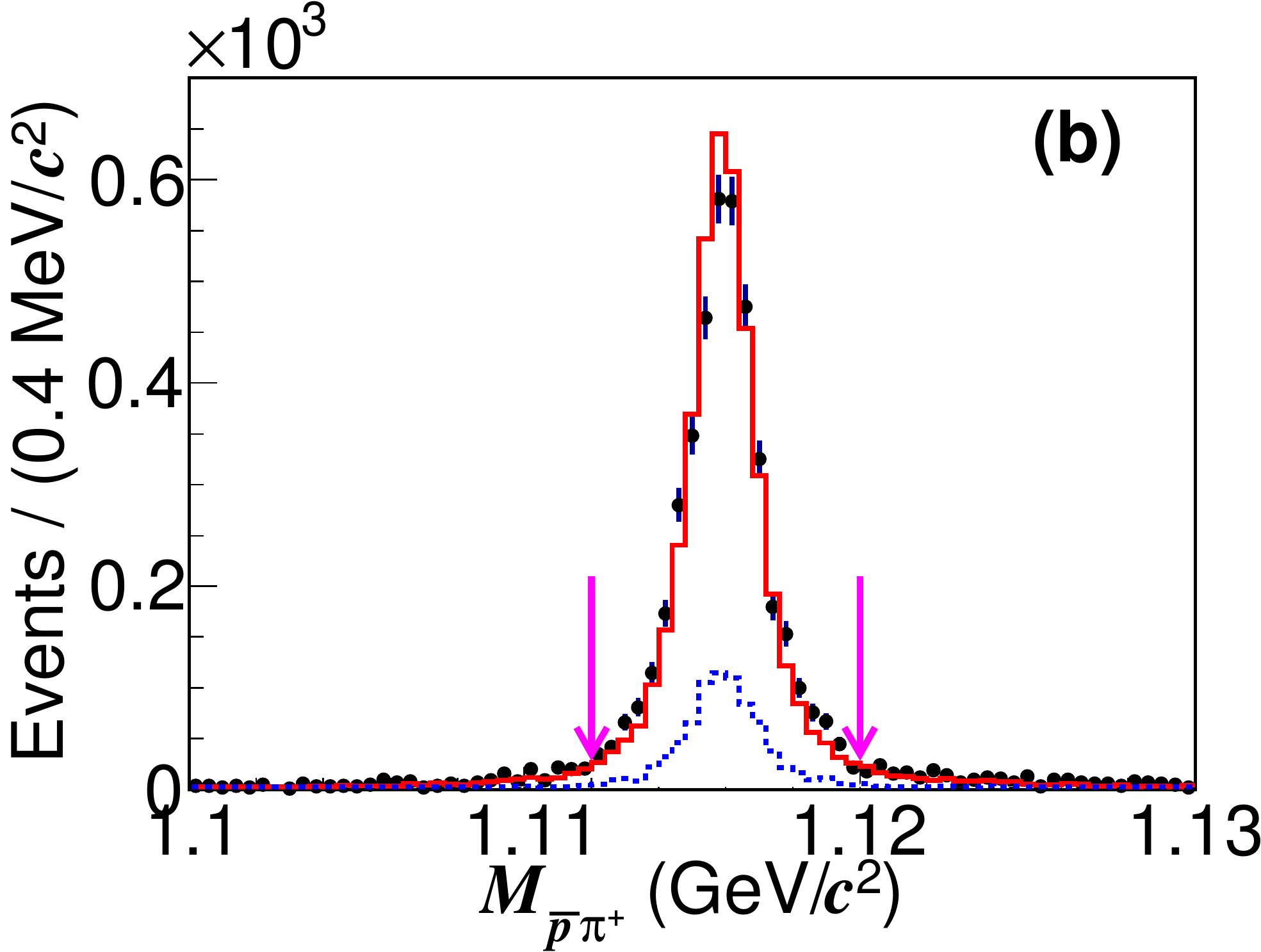}
\includegraphics[width=0.32\textwidth]{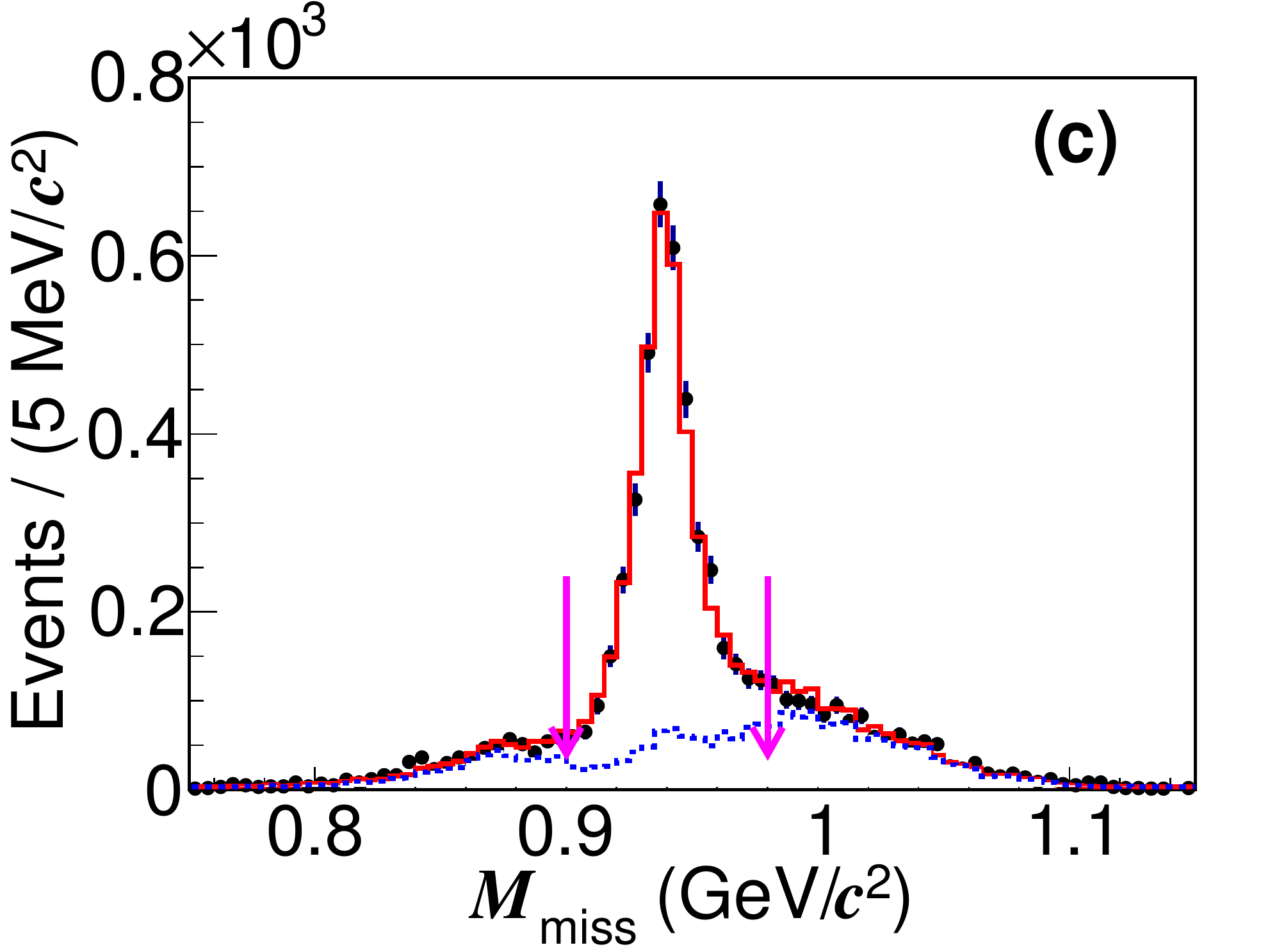}
\caption{
Distributions of (a) $M_{\pi\pi}$ of the $K^0_S$ candidates, (b)
  $M_{\bar p\pi^+}$ of the $\bar\Lambda$ candidates, and (c) $M_{\rm miss}$
  before the 1-C kinematic fit. Dots with error bars are data. Red solid (blue dashed)
  lines refer to the inclusive MC samples with (without) the signal processes.
  The pairs of pink solid (blue dashed) arrows indicate the signal (sideband) regions.
}
\label{fig::ks0}
\end{center}
\end{figure}

Since $K_S^0$ and $\bar\Lambda$ have relatively long lifetimes, they
are reconstructed by constraining the $\pi^{+}\pi^{-}$ pair and the
$\bar{p}\pi^{+}$ pair to originate from secondary vertices,
respectively. Charged track candidates, except the one used as a
$\bar{p}$ in the $\bar\Lambda$ reconstruction, are assumed to be pions
without applying particle identification.  The decay lengths from the
secondary vertex fits of $K_{S}^{0}$ and $\bar\Lambda$ divided by
their corresponding uncertainties are required to be larger than two.
The mass distributions of the reconstructed $K_{S}^{0}$ (denoted as
$M_{\pi\pi}$) and $\bar\Lambda$ (denoted as $M_{\bar{p}\pi^+}$)
candidates are shown in Figs.~\ref{fig::ks0}(a) and~\ref{fig::ks0}(b),
respectively, where the signal regions are defined as $[0.480,
  0.516]$~GeV/$c^{2}$ for $K_{S}^{0}$ and $[1.112, 1.120]$~GeV/$c^{2}$
for $\bar\Lambda$.

Photons are reconstructed as energy clusters in the EMC. The shower
time is required to be within $[0, 700]$~ns from the event start
time. Photon candidates within $|\!\cos\theta| < 0.80$ (barrel) are
required to have deposited energies larger than $25$~MeV and those
with $0.86<|\!\cos\theta|<0.92$~(end cap) must have deposited energies
larger than $50$~MeV.  The photon candidates must be at least
$10\degree$ away from any charged track to suppress Bremsstrahlung
photons or splitoffs.
We require there is at least one photon candidate satisfying the above criteria.

To select the best combination and to improve the resolution, a 1-C
kinematic fit is applied under the hypothesis of $\psi(3686)\to\gamma
nK^0_S\bar\Lambda$, where the neutron is treated as a missing
particle.  The value of $\chi^2_{\rm 1C}$ is required to be less than
200. If more than one combination survives in an event, the one with
the smallest $\chi^2_{\rm 1C}$ is retained.

$M_{\rm miss}$ is defined as the invariant mass of the four
momentum of the missing particle $p_{\rm miss} =p_{\rm CM}- p_{\gamma}- p_{K^0_S} - p_{\Lambda}$,
where $p_{i}$ is the four momentum of the
particle $i$ and $p_{\rm CM}$ is the four momentum of the initial
$e^+e^-$ system. To further improve the purity, $M_{\rm miss}$ before
the 1-C kinematic fit is required to satisfy
$0.90< M_{\rm  miss}<0.98$~GeV/$c^{2}$.  The distribution of
$M_{\rm miss}$ before
the 1-C kinematic fit is shown in Fig.~\ref{fig::ks0}(c).

By analyzing the $\psi(3686)$ inclusive MC samples with
TopoAna~\cite{TopoAna}, the only significant peaking background is
found to be caused by the decays
$\chi_{cJ}\to \Sigma^{\pm}\bar{\Lambda}\pi^{\mp}+c.c.$ with
$\Sigma^{\pm} \to n\pi^{\pm}$. The two pions in this decay may accidentally fall into
the $K_{S}^{0}$ mass window and fulfill the constraint for the
secondary vertex fit, thus faking the $K_{S}^{0}$. This $\Sigma^{\pm}$
background peaks in the invariant mass spectrum of $nK^0_S\bar\Lambda$
(denoted as $M_{nK^0_S\bar{\Lambda}}$) but distributes uniformly in
the $M_{\pi\pi}$ spectrum.  Other backgrounds are smoothly distributed
underneath the $\chi_{cJ}$ signal.

\section{Branching fraction measurement}
\label{chap:BFMEASUREMENT}
\begin{figure}[htbp]
  \begin{center}
    \includegraphics[width=0.48\textwidth]{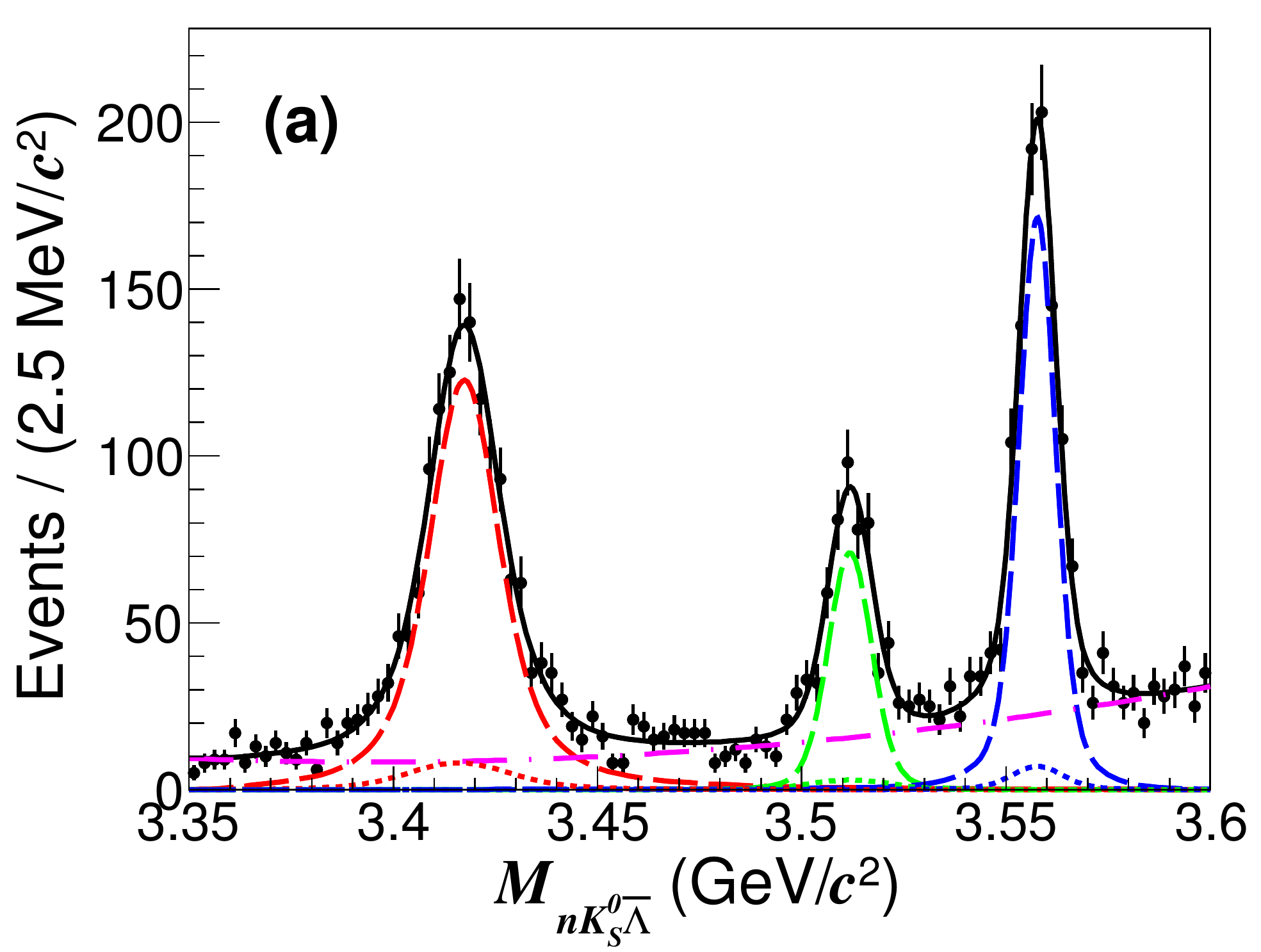}
    \includegraphics[width=0.48\textwidth]{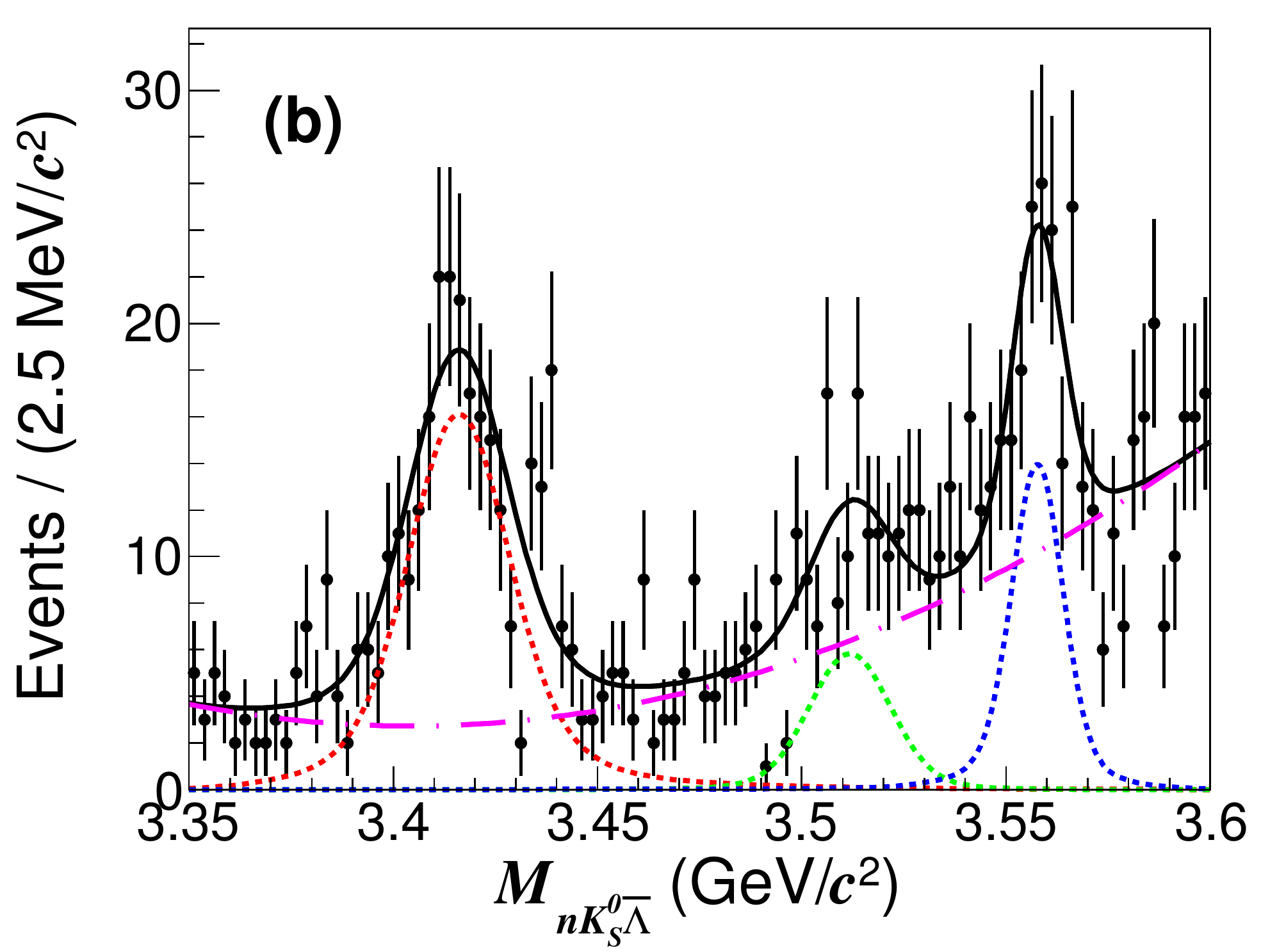}
    \caption{Simultaneous fit to the $M_{nK_{S}^{0}\bar{\Lambda}}$
      spectra in the (a) signal and (b) sideband regions. The points
      with error bars are data. The black solid lines represent the
      total fit. The red, green, and blue dashed lines represent the
      signals of $\chi_{c0}$, $\chi_{c1}$, and $\chi_{c2}$,
      respectively, and the corresponding dotted lines illustrate the
      peaking backgrounds. The purple dotted-dashed lines show the
      fitted backgrounds.
      }
    \label{fig::signalfit}
  \end{center}
\end{figure}

A simultaneous unbinned maximum-likelihood fit to the
$M_{nK^0_S\bar{\Lambda}}$ spectra in both the $M_{\pi\pi}$ signal and
sideband regions, as shown in Fig.~\ref{fig::signalfit}, is performed
to determine the signal yields and peaking backgrounds. The lower and
upper sideband regions are defined as $[0.430, 0.466]$ and $[0.530,
  0.566]$~GeV/$c^{2}$, respectively. The fit model for the signal
region is
\begin{equation}
\sum_{J}(N_{1,J}\cdot f^{J}_{\rm signal}+N_{2,J}\cdot f^{J}_{\rm peakbkg})+N_{3}\cdot f_{\rm flatbkg}\,\, (J= 0,1,2)\,,
\end{equation}
 and that for the sideband region is
\begin{equation}
\sum_{J}(N^{\prime}_{2,J}\cdot f^{J}_{\rm peakbkg})+N^{\prime}_{3}\cdot f^{\prime}_{\rm flatbkg}\,\, (J= 0,1,2)\,.
\end{equation}
The signal shape $f^{J}_{\rm signal}$ for each $\chi_{cJ}$ resonance
is described by its line shape convolved with a double-Gaussian
function to account for the mass resolution. Each signal line shape is
modeled with $BW(M_{nK^0_S\bar{\Lambda}})\times E_{\gamma}^{3}\times
D(E_{\gamma})$, where
$BW(M_{nK^0_S\bar{\Lambda}})=((M_{nK^0_S\bar{\Lambda}}-m_{\chi_{cJ}})^2+0.25\Gamma^2_{\chi_{cJ}})^{-1}$
is the nonrelativistic Breit-Wigner function with the width
$\Gamma_{\chi_{cJ}}$ and the mass $m_{\chi_{cJ}}$ of the corresponding
$\chi_{cJ}$ fixed to the PDG values~\cite{ref::pdg2014},
$E_{\gamma}=(m_{\psi(3686)}^{2}-M_{nK^0_S\bar{\Lambda}}^{2})/2m_{\psi(3686)}$
is the energy of the transition photon in the rest frame of
$\psi(3686)$, and $D(E_{\gamma})$ is a damping factor that suppresses
the divergent tail due to $E_{\gamma}^{3}$. This damping factor is
described by $D(E_{\gamma})={\rm exp}(-E_{\gamma}^{2}/8\beta^{2})$,
where $\beta=(65.0\pm 2.5)$~MeV is measured by the CLEO
collaboration~\cite{ref::beta}.  The two Gaussian functions in the
convolution share the same mean value which is then floated in the
fit. The relative width and size of the second Gaussian to the first
Gaussian function are fixed to the results of MC studies, while the
width of the first Gaussian function is floated. The peaking
background shapes $f^{J}_{\rm peakbkg}$ are parameterized the same as the
signal shapes.  The yields $N_{2,J}$ in the signal region are
normalized to $N^{\prime}_{2,J}$ in the $M_{\pi\pi}$ sideband
regions according to the sizes of these two regions. The background shapes
$f^{(\prime)}_{\rm flatbkg}$ in both
regions are modeled as second-order Chebyshev polynomial functions. The
numbers of fitted $\chi_{cJ}$ signal events, $N_{1,J}$, are listed in
Table~\ref{tab::results}.

A special generator based on results of HelPWA~\cite{ref::chung,ref::blatt,ref::helb3} for the decay
$\chi_{cJ}\to nK^0\bar\Lambda+c.c.$ is developed to estimate the
detection efficiencies.
The description of HelPWA can be found in the supplemental material~\cite{SM}.
For the $\chi_{cJ}$ data events used in HelPWA, the signal regions of $M_{nK_{S}^{0}\bar{\Lambda}}$ for $\chi_{c0}$,
  $\chi_{c1}$, and $\chi_{c2}$ are $[3.39, 3.45]$, $[3.50, 3.53]$, and
$[3.54, 3.57]$~GeV/$c^{2}$, respectively; the masses of $K^0$, $\bar\Lambda$, and $\chi_{cJ}$ are
constrained to their known masses~\cite{ref::pdg2014}. The inclusive background MC
sample is used to calculate the background likelihood with negative weight.

\begin{figure}[htbp]
\begin{center}
\includegraphics[width=0.32\textwidth]{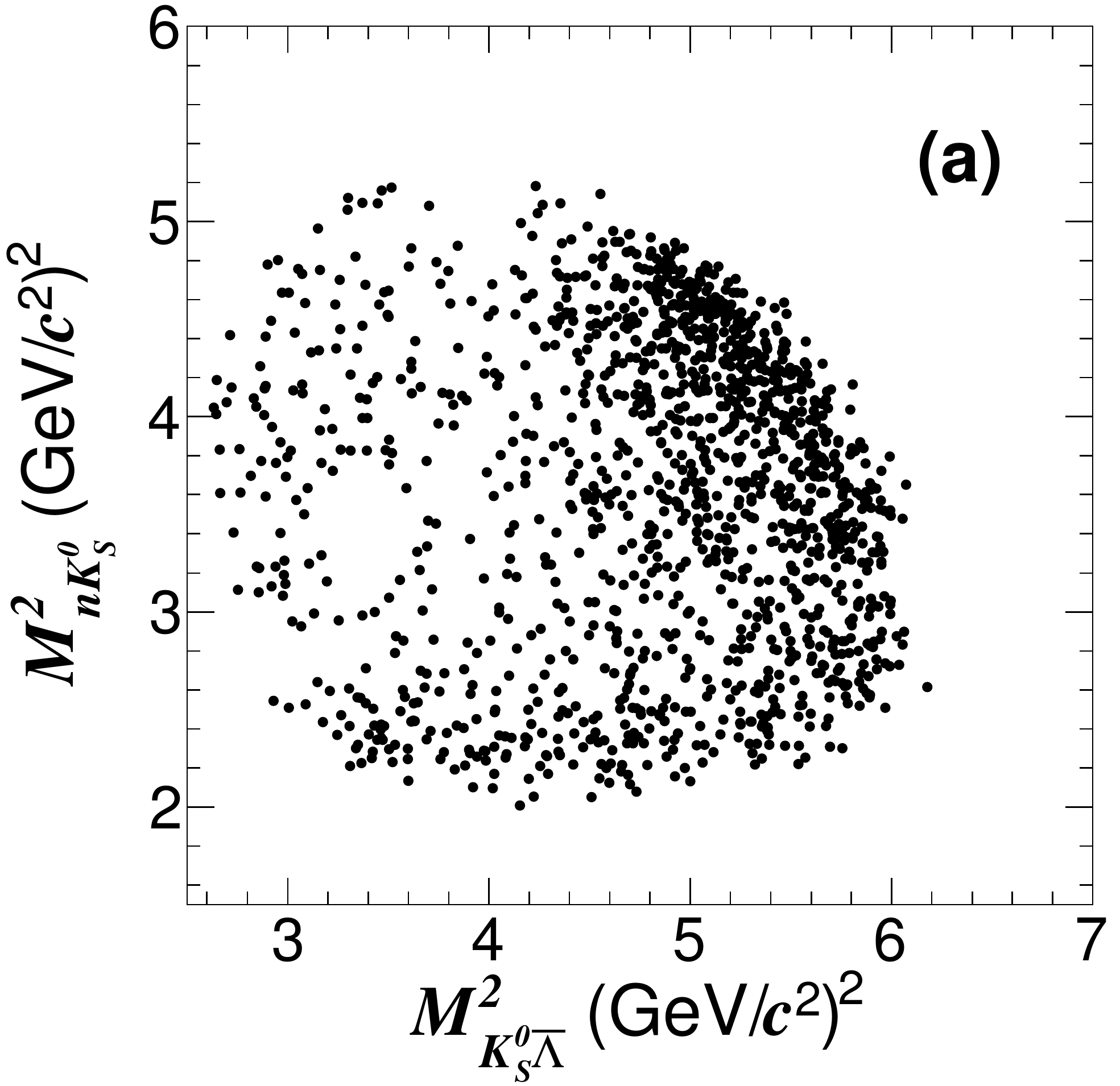}
\includegraphics[width=0.32\textwidth]{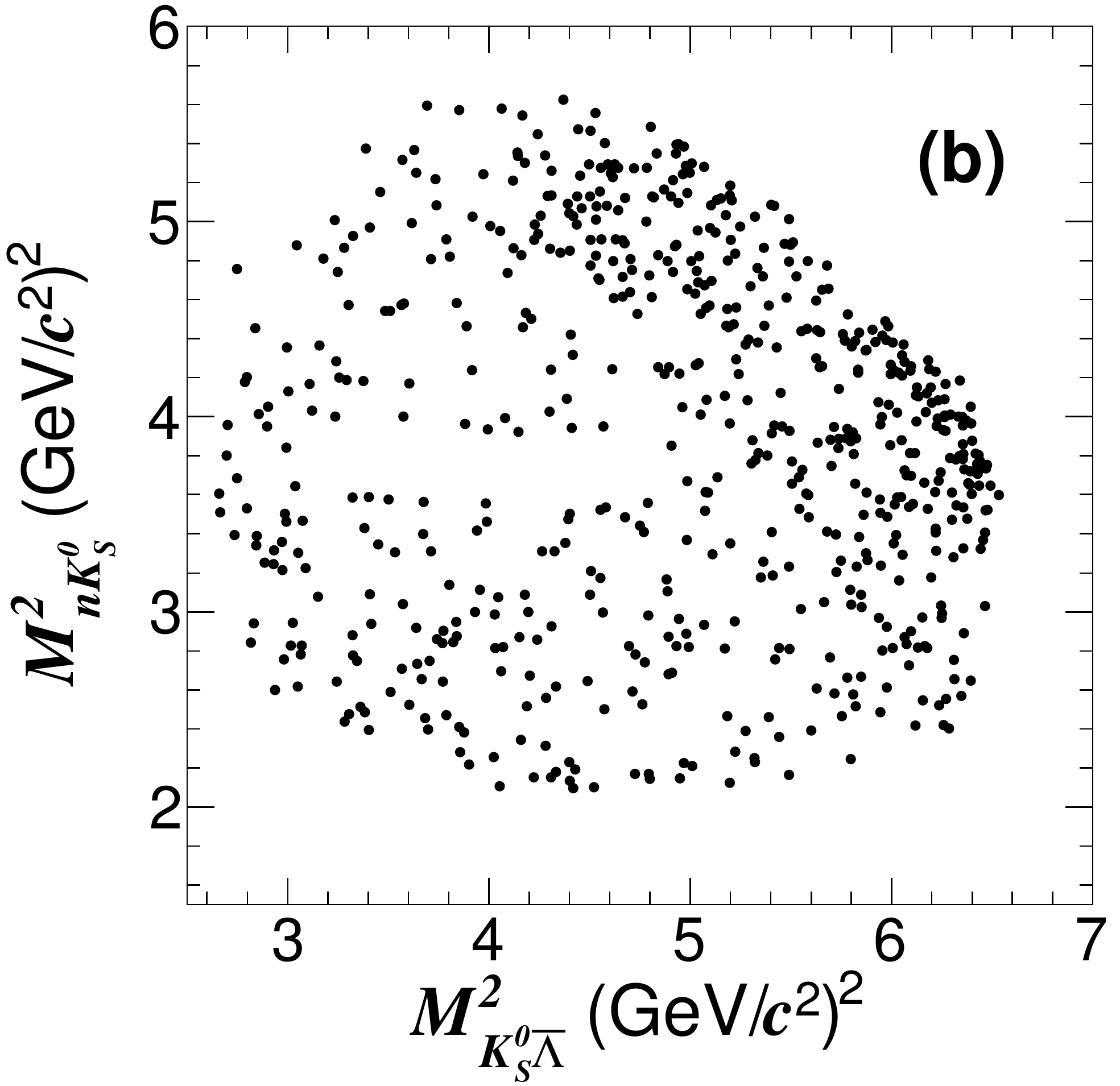}
\includegraphics[width=0.32\textwidth]{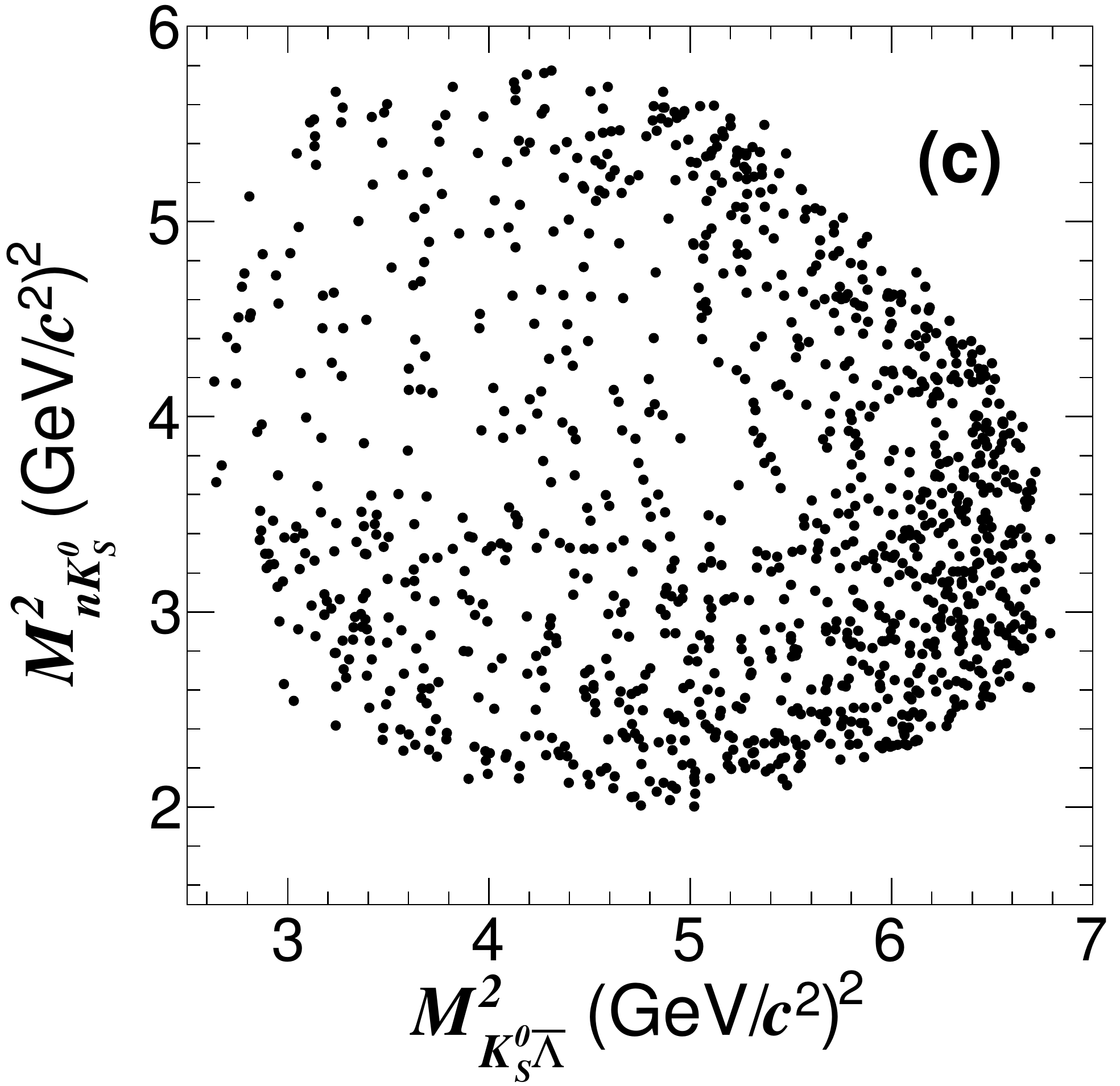}
\caption{Dalitz plots of $M^2_{nK^0_S}$ versus
  $M^2_{K^0_S\bar{\Lambda}}$ for the (a) $\chi_{c0}$, (b) $\chi_{c1}$,
  and (c) $\chi_{c2}$ candidates in the data sample.  The signal
  regions of $M_{nK_{S}^{0}\bar{\Lambda}}$ for $\chi_{c0}$,
  $\chi_{c1}$, and $\chi_{c2}$ are $[3.39, 3.45]$, $[3.50, 3.53]$, and
  $[3.54, 3.57]$~GeV/$c^{2}$, respectively.  } \label{fig::dalitz}
\end{center}
\end{figure}

\begin{figure}[htbp]
  \begin{center}	
      \includegraphics[width=0.32\textwidth]{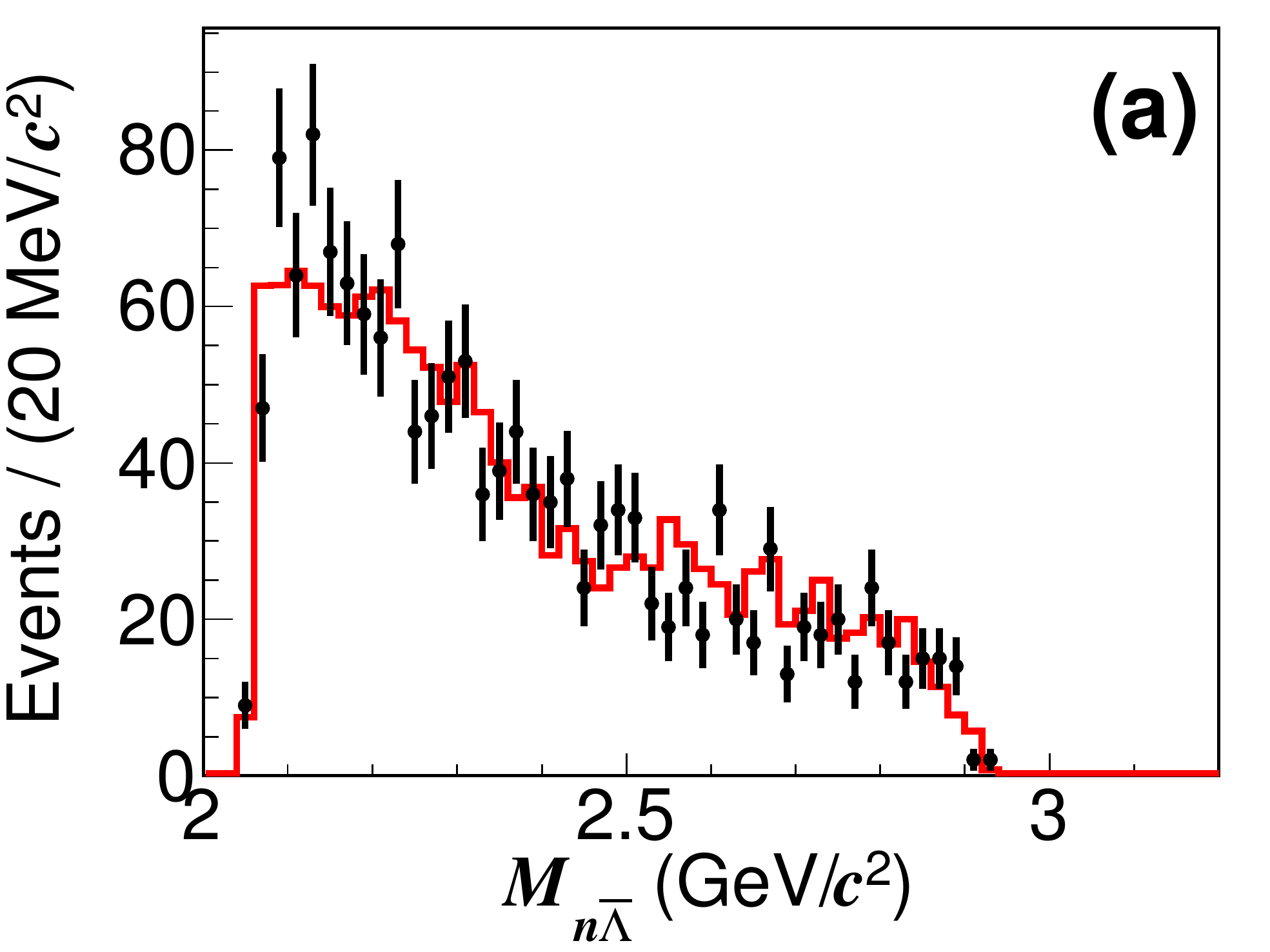}
      \includegraphics[width=0.32\textwidth]{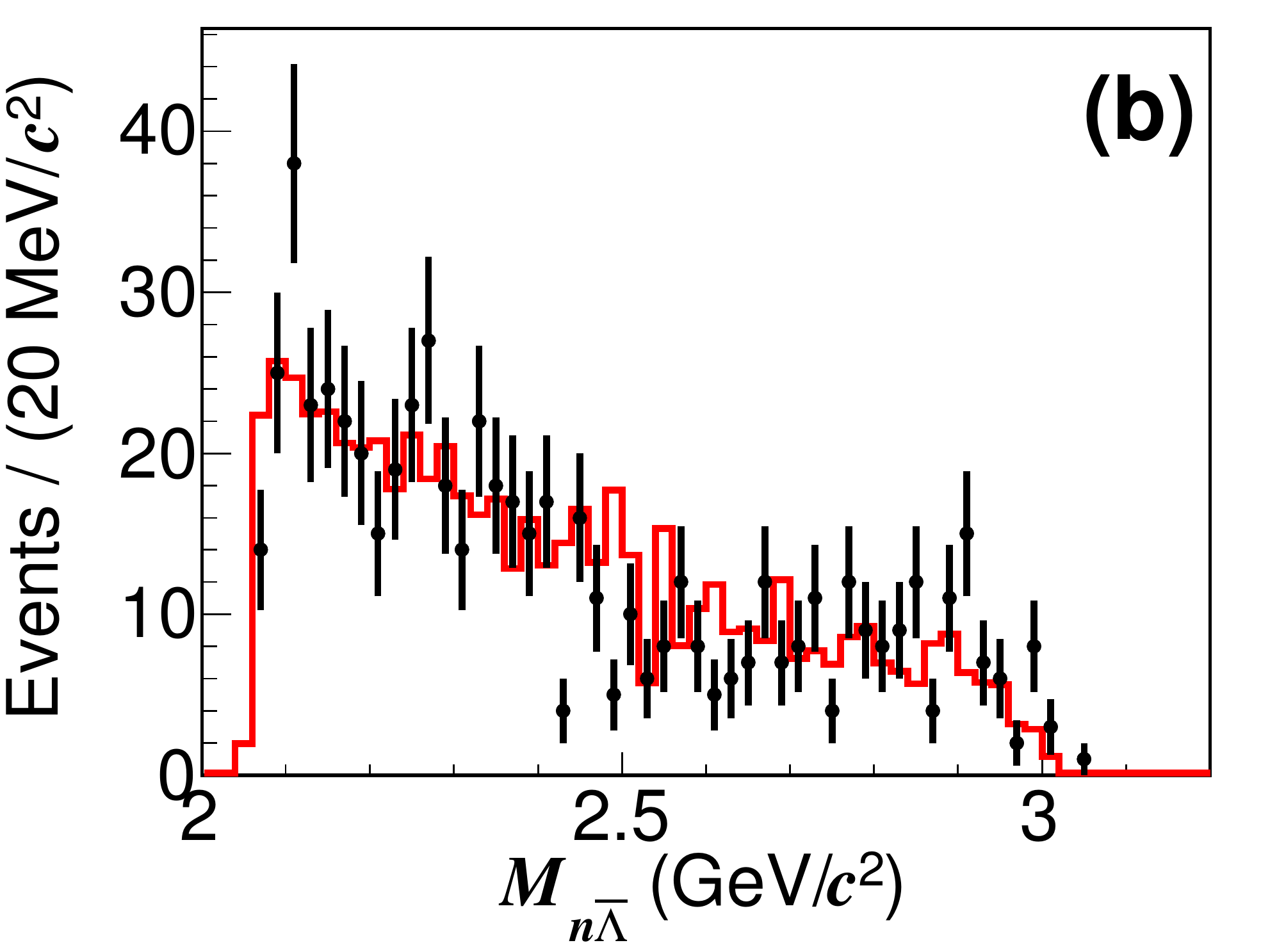}
      \includegraphics[width=0.32\textwidth]{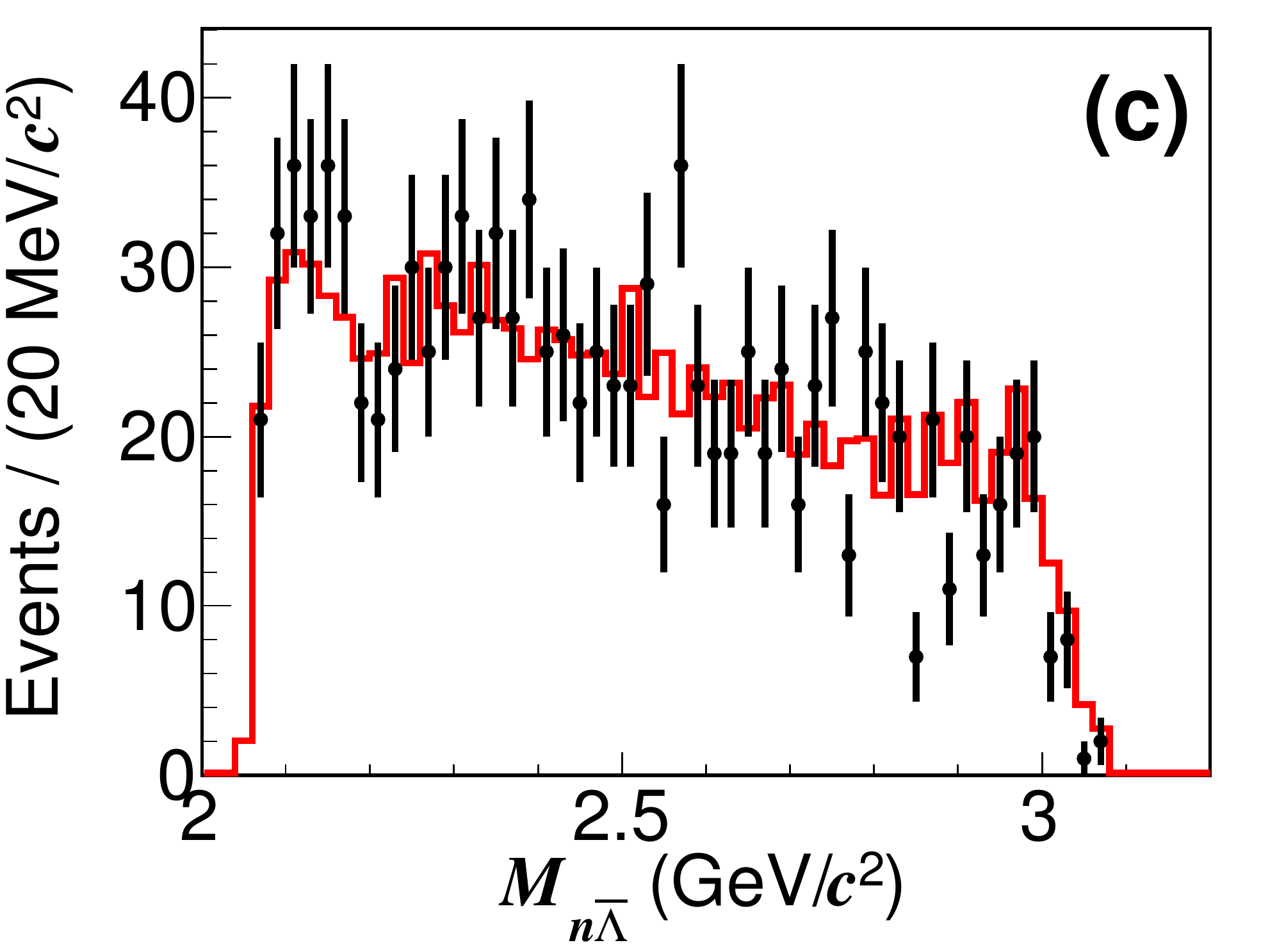}\\
      \includegraphics[width=0.32\textwidth]{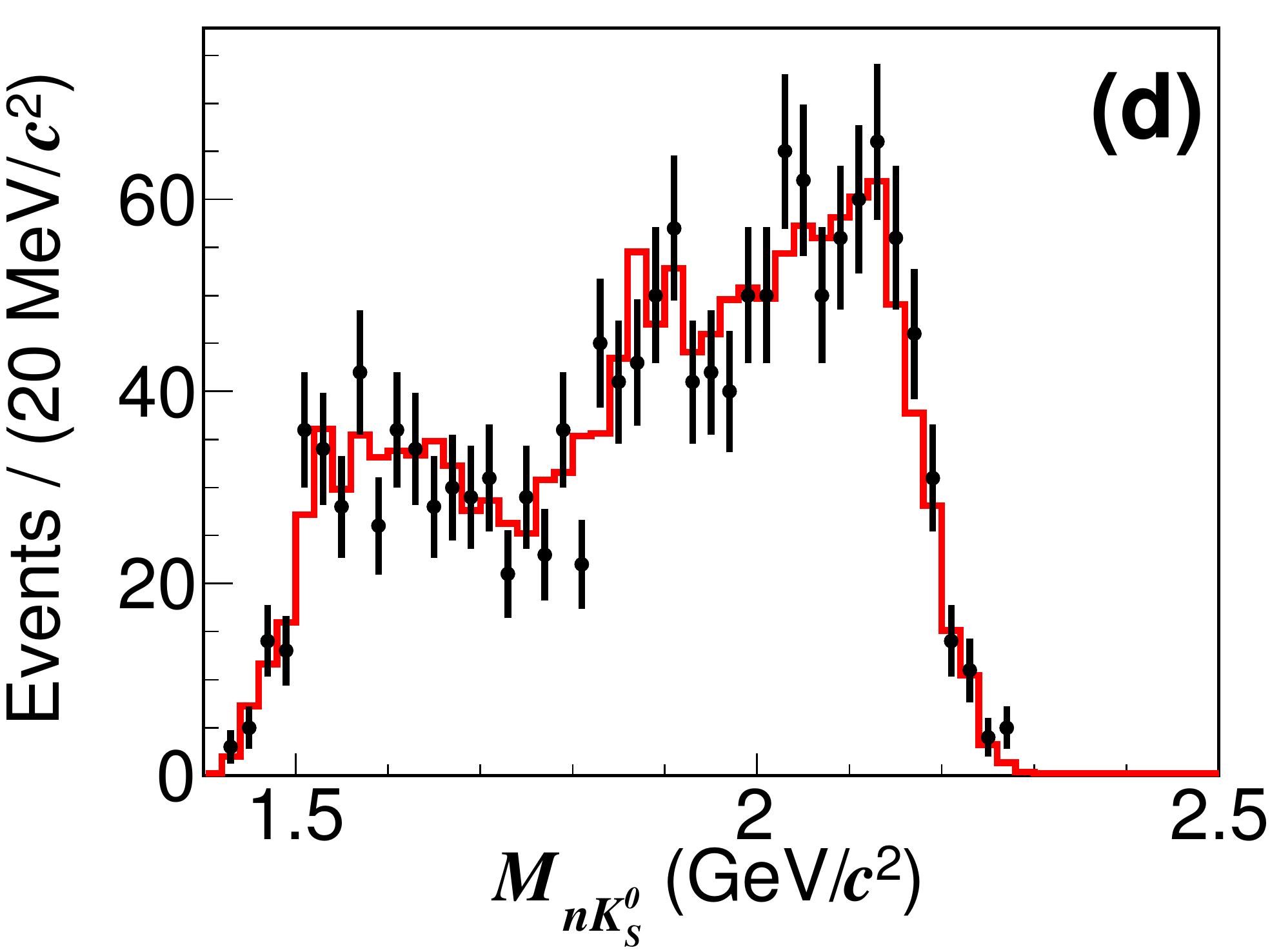}
      \includegraphics[width=0.32\textwidth]{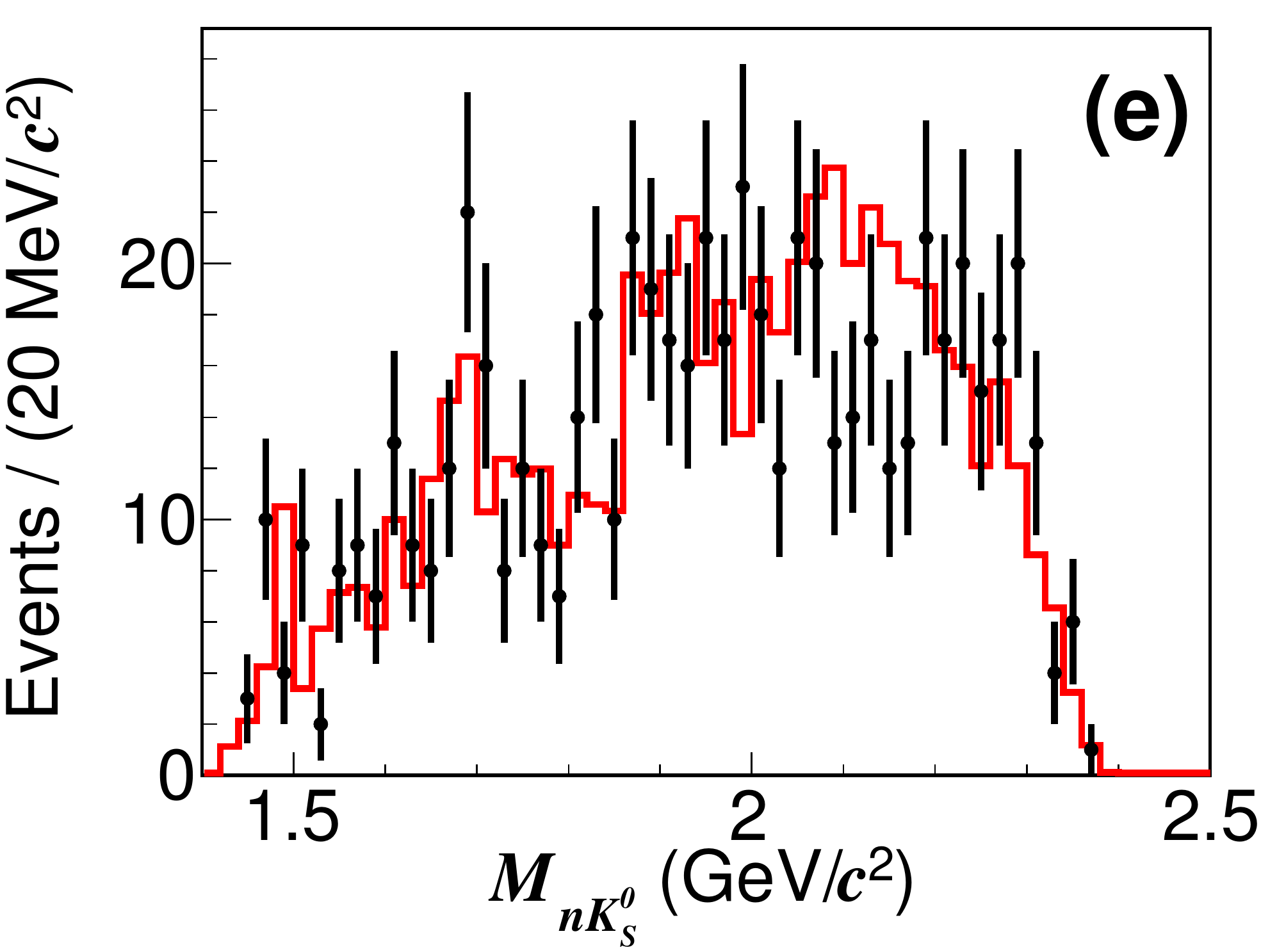}
      \includegraphics[width=0.32\textwidth]{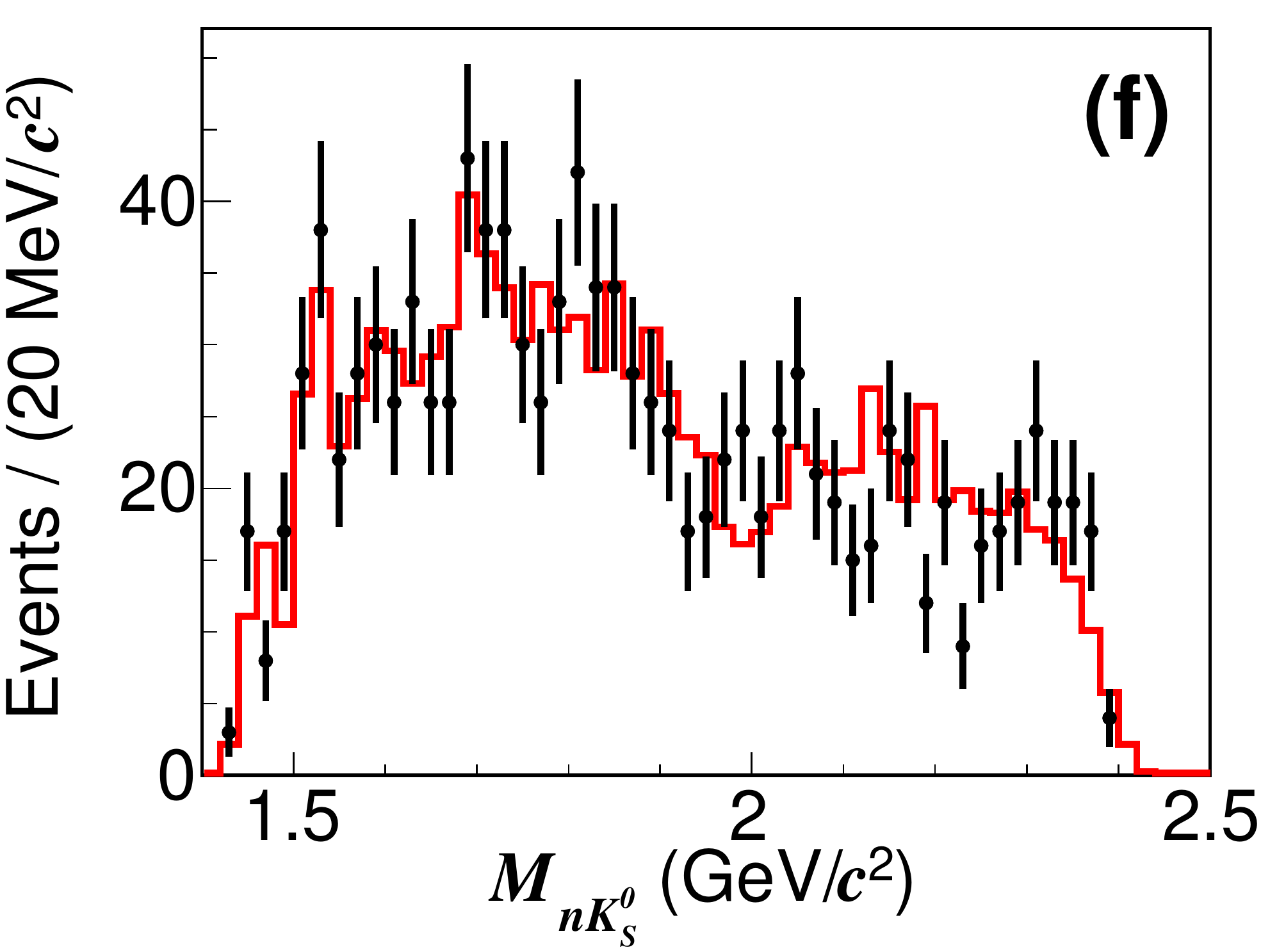}\\
      \includegraphics[width=0.32\textwidth]{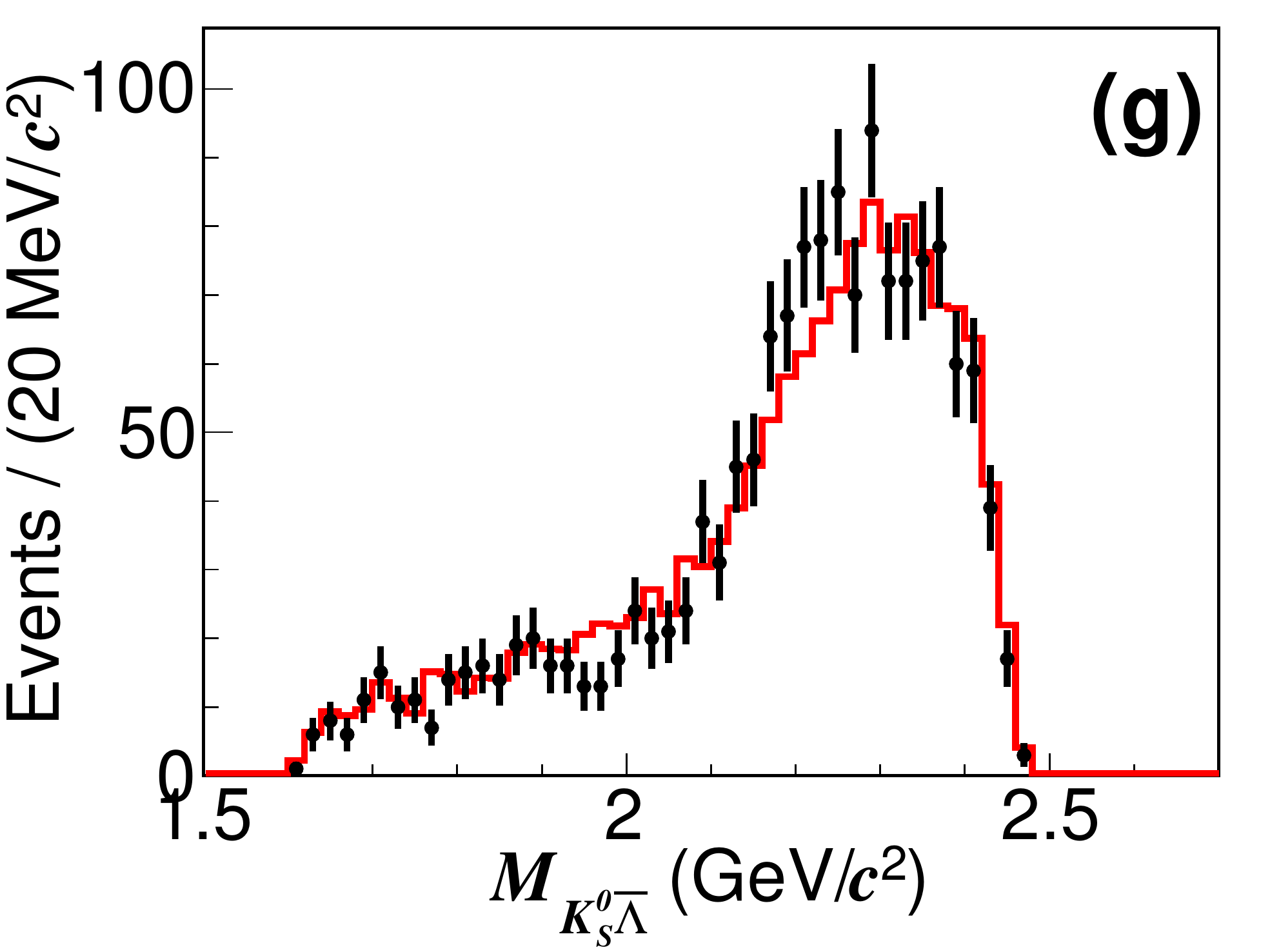}
      \includegraphics[width=0.32\textwidth]{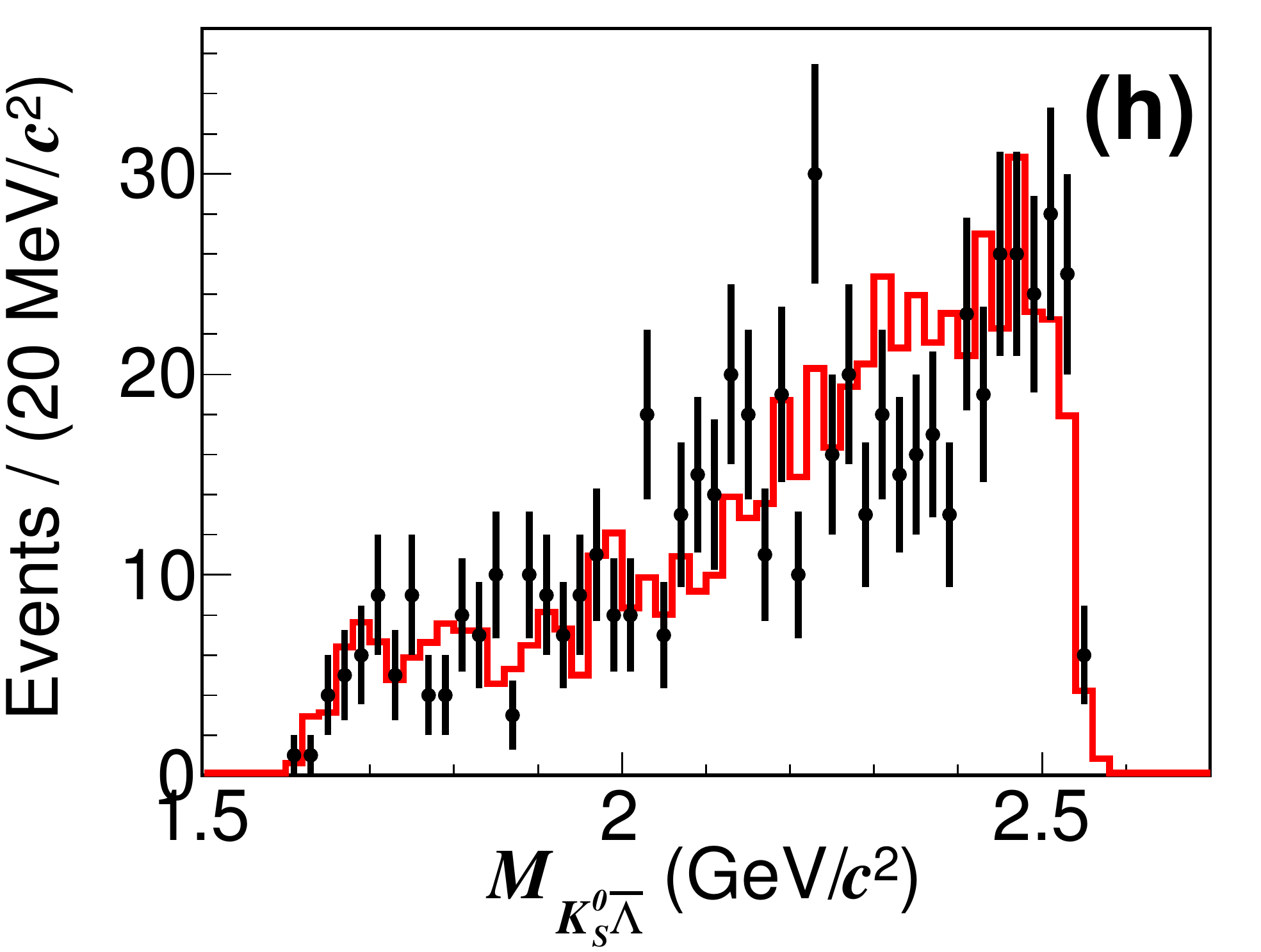}
      \includegraphics[width=0.32\textwidth]{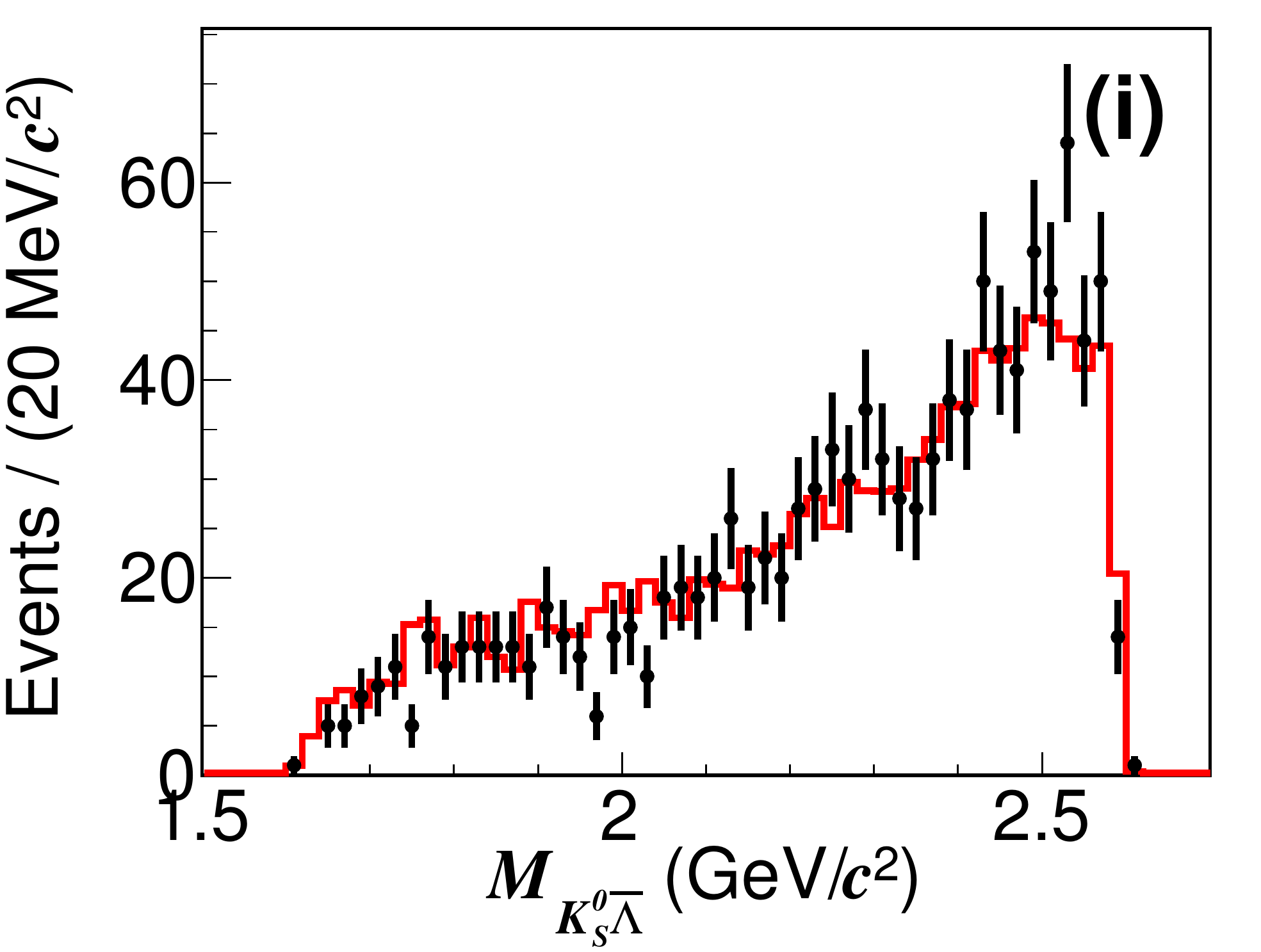}\\
    \caption{Distributions of $M_{n\bar{\Lambda}}$, $M_{nK^0_S}$, and $M_{K^0_S\bar{\Lambda}}$
       for $\chi_{cJ}\to nK^0_S\bar\Lambda+c.c.$ ($J=0$,~1,~2).
      The left column is for $\chi_{c0}$, the middle column
      is for $\chi_{c1}$, and the right column is
      for $\chi_{c2}$. The data are represented by black points with error
      bars and the MC events are represented by red lines.
      The signal regions of $M_{nK_{S}^{0}\bar{\Lambda}}$ for $\chi_{c0}$, $\chi_{c1}$, and $\chi_{c2}$ are
      $[3.39, 3.45]$, $[3.50, 3.53]$, and $[3.54, 3.57]$~GeV/$c^{2}$, respectively.
    } \label{fig::masses-helpwa}
  \end{center}
\end{figure}

The signal MC events are generated with the HelPWA model, in which the parameters of coupling constants are
determined by fitting the model to the $\chi_{cJ}$ data events.
The Dalitz plots and the two-body invariant mass $M_{ij}$ distributions
of the data sample are shown in Figs.~\ref{fig::dalitz} and~\ref{fig::masses-helpwa},
respectively, where $i$ and $j$ denote the final particles. The
generated signal MC events based on HelPWA along with the simulated
background events from the inclusive MC sample are represented as the solid
lines in Fig.~\ref{fig::masses-helpwa}.
The signal MC samples are generated with $\chi_{cJ}\to nK_{S}^{0}\bar{\Lambda}$ and
$\chi_{cJ}\to \bar{n}K_{S}^{0}\Lambda$ separately. After applying the
same event selection criteria to the signal MC samples, we fit the
invariant mass spectra with the same methods used for the experimental
data. The detection efficiencies of $\chi_{cJ}$, $\epsilon_{J}$, are
averaged over both charge conjugate channels and listed in
Table~\ref{tab::results}.  The efficiency differences between the two
charged conjugated modes are less than 0.5\% for all three $\chi_{cJ}$
channels and are consistent within the statistical uncertainties.

The BFs for $\chi_{cJ}\to nK^0_S\bar\Lambda$ are calculated using
\begin{equation} \mathcal{B}(\chi_{cJ}\to
nK^0_S\bar\Lambda)=\frac{N_{1,J}}{N_{\psi(3686)}\cdot\epsilon_{J}\cdot\prod_{i}\mathcal{B}_{i}}\,,\label{eq:BF}
\end{equation} where $N_{\psi(3686)}$ is the number of
$\psi(3686)$ events~\cite{ref::CPC42_023001}; $\epsilon_{J}$ is
the detection efficiency as listed in
Table~\ref{tab::results};
$\prod_{i}\mathcal{B}_{i}=\mathcal{B}(\psi(3686)\to\gamma\chi_{cJ})\cdot\mathcal{B}(K^0_S\to\pi^+\pi^-)\cdot\mathcal{B}(\bar\Lambda\to
\bar{p}\pi^+)$, where the BFs are taken from the
PDG~\cite{ref::pdg2014}.  The results are summarized in
Table~\ref{tab::results}.

\begin{table}[htbp]
\begin{center}
\caption{The number of fitted signal events~($N_{1,J}$), detection
  efficiency~($\epsilon_{J}$), and $\mathcal{B}(\chi_{cJ}\to nK^0_S\bar\Lambda+c.c.)$,
  where the first uncertainty is statistical and the second one is systematic.
}\label{tab::results}
\begin{tabular}{lccc}
\hline
\hline
Mode         & $N_{1,J}$     & $\epsilon_{J}$ (\%) & BF ($10^{-4}$) \\
\hline
$\chi_{c0}$  & $1288 \pm 50$ & 9.95            & $6.67 \pm 0.26 \pm 0.41$ \\
$\chi_{c1}$  & $410 \pm 30$  & 12.44           & $1.71 \pm 0.12 \pm 0.12$ \\
$\chi_{c2}$  & $900 \pm 41$  & 13.03           & $3.66 \pm 0.17 \pm 0.23$ \\
\hline
\hline
\end{tabular}
\end{center}
\end{table}

\section{Systematic uncertainty} \label{chap:SYSTEMATICS}
The number of $\psi(3686)$ events is measured to be $(4.48\pm0.03)\times10^8$
based on inclusive hadronic events, as described in
Ref.~\cite{ref::CPC42_023001}, so the uncertainty is 0.6\%.
The systematic uncertainty due to the detection of $\gamma$ is studied with the
well understood channel $J/\psi\to\rho^0\pi^0$~\cite{ref::gamma-recon}. The
efficiency difference between data and MC simulation is about 1\% per photon.
To estimate the uncertainties associated with $K^0_S$ and $\Lambda$
reconstruction, the decays $J/\psi\to K^{*\pm}(892)K^{\mp}$,
$K^{*\pm}(892)\to K^0_S\pi^{\pm}$ and $J/\psi\to\Lambda\bar\Lambda$ are
selected as the control samples. The uncertainties are determined to be $1.5\%$
per $K^0_S$ and $2.0\%$ per $\Lambda$.
The systematic uncertainty caused by the 1-C kinematic fit is studied with the
control sample $\psi(3686)\to\pi^0nK^0_S\bar\Lambda$ with purity about 98.5\%,
where $n$, $K^0_S$, and $\bar\Lambda$ are selected using the same criteria as
the nominal analysis but the number of good photons is required to be at least
two. The difference of the selection efficiencies between data and MC
simulation is determined to be $4.1\%$ and assigned as the corresponding
systematic uncertainty.
The uncertainties associated with the mass windows of $K_{S}^{0}$, $\Lambda$,
and $M_{\rm miss}$ are estimated by repeating the analysis with alternative
mass window requirements. We change the mass window of $K_{S}^{0}$ to
$[0.473, 0.521]$ and $[0.487, 0.511]$~GeV/$c^{2}$, that of $\Lambda$ to
$[1.110, 1,123]$ and $[1.113, 1.119]$~GeV/$c^{2}$, and that of $M_{\rm miss}$
to [0.880, 1.000] and [0.910, 0.970]~GeV/$c^{2}$.  The largest differences from
the nominal BFs are assigned as the corresponding systematic uncertainties.

The systematic uncertainty related to the fitting procedure includes
multiple sources. For the signal line shape, the parameterization of
the damping factor may introduce a systematic uncertainty. The nominal
damping factor is changed to another popular form used by
KEDR~\cite{ref::damping}, given by $D(E_{\gamma})=\frac{E_{\gamma
0}^2}{E_{\gamma}E_{\gamma 0}+(E_{\gamma}-E_{\gamma 0})^2}$, where
$E_{\gamma 0}=(m_{\psi(3686)}^{2}-M_{\chi_{cJ}}^{2})/2m_{\psi(3686)}$.
The resulting differences in the fit are assigned as the related
systematic uncertainties.  In addition, the background function is
changed from a second to a third order Chebyshev function, and the
differences in signal yields are taken as the systematic
uncertainties. The systematic uncertainty due to the fit range is
evaluated by changing the fit range from $[3.35, 3.60]$ to $[3.35,
3.65]$ and $[3.30, 3.65]$~GeV/$c^{2}$, and the maximum differences in
the fitted yields are considered as the associated systematic
uncertainties.  The total uncertainties of the fitting procedure are
estimated to be 2.8\%, 4.1\%, and 2.3\% for $\chi_{c0}$, $\chi_{c1}$,
and $\chi_{c2}$, respectively.

The systematic uncertainties arising from MC modeling are estimated by
using different model parameters and some unimportant
intermediate processes in HelPWA.  The differences of efficiencies
based on the new HelPWA results and the nominal ones are taken as the
uncertainties.  The systematic uncertainties due to the input BFs of
$\psi(3686)\to\gamma\chi_{c0}$ ($\chi_{c1}$, $\chi_{c2}$),
$K^0_S\to\pi^+\pi^-$, and $\Lambda\to p\pi^-$ are 2.0\% (2.5\%,
2.1\%), 0.1\%, and 0.8\%, respectively, according to the
PDG~\cite{ref::pdg2014}.

All systematic uncertainty contributions are summarized in
Table~\ref{tab::sys}. The total systematic uncertainty for each $\chi_{cJ}$
decay is obtained by adding all contributions in quadrature.

\begin{table}[htbp]
  \begin{center}
    \caption{Systematic uncertainty sources and their
      contributions~(in \%).}\label{tab::sys}
    \begin{tabular}{lccc}
      \hline
      \hline
      Sources & $\mathcal{B}(\chi_{c0})$ & $\mathcal{B}(\chi_{c1})$ &  $\mathcal{B}(\chi_{c2})$ \\
      \hline
      $N_{\psi(3686)}$                             & 0.6 & 0.6 & 0.6 \\
      $\gamma$ detection                           & 1.0 & 1.0 & 1.0 \\
      $K^0_S$ reconstruction                       & 1.5 & 1.5 & 1.5 \\
      $\Lambda$ reconstruction                     & 2.0 & 2.0 & 2.0 \\
      Kinematic fit                                & 4.1 & 4.1 & 4.1 \\
      Mass windows                                 & 0.4 & 0.7 & 0.5 \\
      Fitting procedure                            & 2.8 & 4.1 & 2.3 \\
      MC modeling                                  & 1.3 & 1.3 & 2.3 \\
      Input BFs                                    & 2.2 & 2.6 & 2.2 \\
      \hline
      Total                                        & 6.2 & 7.1 & 6.3 \\
      \hline
      \hline
    \end{tabular}
  \end{center}
\end{table}

\section{Summary}
\label{chap:SUMMARY}
The decays of $\chi_{cJ}\to nK^0_S\bar\Lambda+c.c.$ ($J=0$,~1,~2) are observed
for the first time using $(4.48\pm0.03)\times 10^{8}$ $\psi(3686)$ events
accumulated with the BESIII detector at the BEPCII collider. The BFs of
$\chi_{cJ}\to nK^0_S\bar\Lambda + c.c.$ are determined to be
$(6.67 \pm 0.26_{\rm stat} \pm 0.41_{\rm syst})\times10^{-4}$,
$(1.71 \pm 0.12_{\rm stat} \pm 0.12_{\rm syst})\times10^{-4}$, and
$(3.66 \pm 0.17_{\rm stat} \pm 0.23_{\rm syst})\times10^{-4}$ for $J=0$, 1, and
2, respectively. Isospin symmetry is examined by comparing our results with the
isospin conjugate decays of $\chi_{cJ}\to pK^-\bar\Lambda+c.c.$~\cite{ref::paper-pkl}.
The ratios
$\mathcal{B}(\chi_{cJ}\to pK^-\bar\Lambda+c.c.)/\mathcal{B}(\chi_{cJ}\to nK^0_S\bar\Lambda+c.c.)$
are measured to be $1.98 \pm 0.09_{\rm stat} \pm 0.14_{\rm syst}$,
$2.64 \pm 0.23_{\rm stat} \pm 0.20_{\rm syst}$, and
$2.29 \pm 0.13_{\rm stat} \pm 0.16_{\rm syst}$ for $J=0$, 1, and 2,
respectively, where common sources of systematic uncertainties are canceled. No
obvious isospin violation is observed.

Enhancements are observed in the Dalitz plots shown in Fig.~\ref{fig::dalitz}
and the mass distributions of two-body $n\bar\Lambda$ subsystems shown in
Fig.~\ref{fig::masses-helpwa}. We perform a HelPWA with the main goal to
produce MC samples that describe the data well enough to obtain a good estimate
of the efficiency. While the HelPWA does describe the data nicely, the
complexity of the model we used here does not allow to draw any firm
conclusions on the relative contributions of individual resonances.

\acknowledgments
The BESIII collaboration thanks the staff of BEPCII and the IHEP computing center for their strong support. This work is supported in part by National Key Research and Development Program of China under Contracts Nos.~2020YFA0406300, 2020YFA0406400; National Natural Science Foundation of China (NSFC) under Contracts Nos.~11875170, ~11875115, ~11475090, ~11625523,~11635010, ~11735014, ~11822506, ~11835012, ~11875054, ~11875262, ~11935015, ~11935016, ~11935018, ~11961141012, ~12022510, ~12035009, ~12035013, ~12061131003; the Chinese Academy of Sciences (CAS) Large-Scale Scientific Facility Program; Joint Large-Scale Scientific Facility Funds of the NSFC and CAS under Contracts Nos. U2032110, U1732263, U1832207, U2032104, U2032110; CAS Key Research Program of Frontier Sciences under Contract No.~QYZDJ-SSW-SLH040; 100 Talents Program of CAS; INPAC and Shanghai Key Laboratory for Particle Physics and Cosmology; ERC under Contract No. 758462; European Union Horizon 2020 research and innovation programme under Contract No. Marie Sklodowska-Curie grant agreement No 894790; German Research Foundation DFG under Contracts No.~443159800, Collaborative Research Center CRC 1044, FOR 2359, FOR 2359, GRK 214; Istituto Nazionale di Fisica Nucleare, Italy; Ministry of Development of Turkey under Contract No. DPT2006K-120470; National Science and Technology fund; Olle Engkvist Foundation under Contract No. 200-0605; STFC (United Kingdom); The Knut and Alice Wallenberg Foundation (Sweden) under Contract No. 2016.0157; The Royal Society, UK under Contracts No.~DH140054, DH160214; The Swedish Research Council; U.~S.~Department of Energy under Contracts Nos.~DE-FG02-05ER41374, DE-SC-0012069.

\bibliographystyle{JHEP}
\bibliography{references.bib}

\end{document}


\maketitle
\flushbottom

\section{HelPWA Model}
\subsection{Variable definition}
\begin{figure}[htbp]
\begin{center}
\includegraphics[width=5in,angle=0]{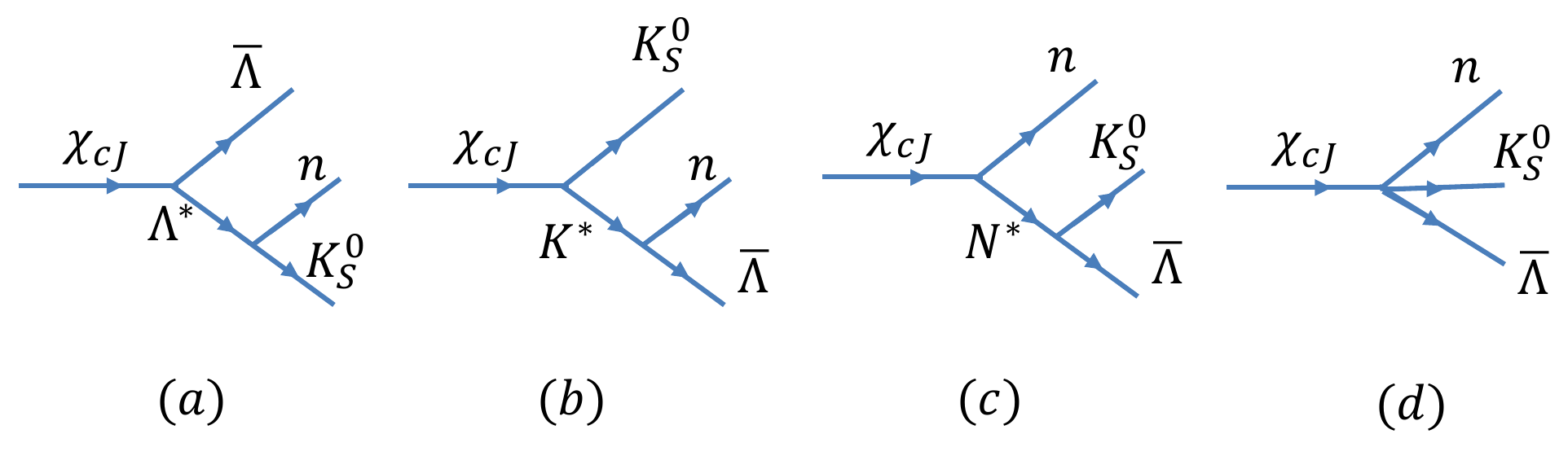}
\put(-280,90){\rotatebox{90}} \caption{The quasi-two body decays and the direct three-body decays in the process $\chicj\to\pp\lmdb$.} \label{fig1}
\end{center}
\end{figure}

Four-momentum vectors of final states for the decay $\chicj\to\pp\lmdb$ are denoted by $p_1,p_2$ and $p_3$. Momentum combinations, $p_i+p_j$ and $p_i+p_j+p_k$, are denoted by $[ij]$ and $[ijk]$, respectively. The production of $\pp\lmdb$ events, as shown in Fig. \ref{fig1}, is assumed via intermediate states $\Lambda^*,~K^*,~N^*$ and direct three-body decay,
\begin{eqnarray}
(a): & \chicj\to\Lambda^*\lmdb, ~\Lambda^*\to\pp,\\
(b): & \chicj\to K_S^0K^{*+}, ~K^{*+}\to n\lmdb,\\
(c): & \chicj\to n\bar N^*, ~\bar N^{*}\to K_S^0\lmdb.\\
(d): & \chicj\to n K_S^0\lmdb.
\end{eqnarray}

\begin{figure}[htbp]
\begin{center}
\includegraphics[width=5in,angle=0]{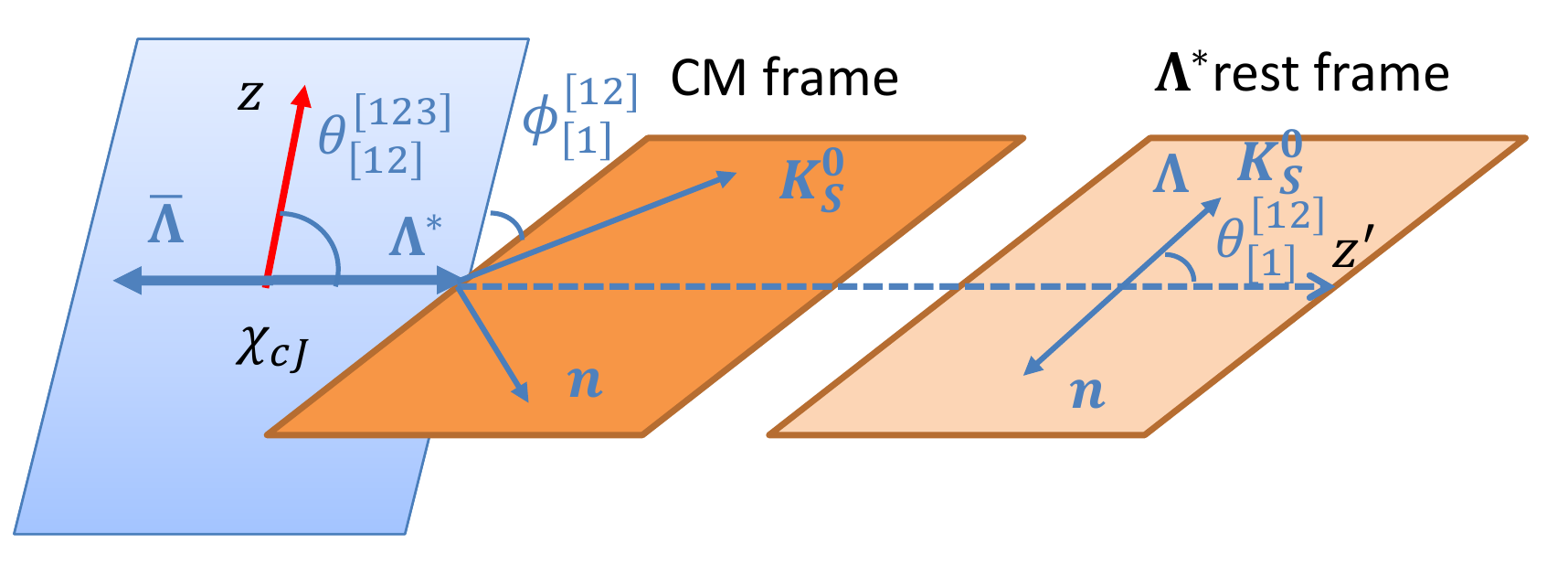}
\put(-280,90){\rotatebox{90}} \caption{Helicity system of the decay $\chicj\to\Lambda^*\lmdb,~\Lambda^*\to nK_S^0$.} \label{helsys}
\end{center}
\end{figure}
For the decay $\chicj\to A(p_A)B(p_B)$, the helicity angles are taken as the polar and azimuthal angles for the momentum ${\bf p}_A$ in the $\chicj$ rest frame. Let $\hat{x},~\hat{y}$ and $\hat{z}$ be the coordinate system fixed in the $\chicj$ rest frame. For the second decay, $A\to X(p_X)Y(p_Y)$, the $z$ axis, $\hat{z}_h$, is taken along the flying direction of the $A$ in the $\chicj$ rest frame, and momenta $p_X$ and $p_Y$ are defined in the $A$ rest frame. Then the helicity coordinate system fixed by the $A$ decay is defined as $\hat{y}_h\propto\hat{z}\times\hat{z}_h$ and $\hat{x}_h=\hat{y}_h\times\hat{z}_h$. The helicity angles for the second decay are taken as the polar and azimuthal angles for the momentum ${\bf p}_X$ in the helicity system ($\hat{x}_h,\hat{y}_h,\hat{z}_h$). Figure \ref{helsys} shows the helicity system definition for type (a) decay, for example. Variable definitions for the helicity angles and amplitudes are given in Table \ref{tab1} for the sequential decays $(a)$, $(b)$ and $(c)$.

\begin{table}[htbp]
\begin{center}
\caption{Helicity angles and amplitudes of the sequential decays (a), (b) and (c). $\lambda_i$ denotes the value of helicity for the corresponding particle, and $m$ denotes the spin $z$ projection of virtual photon $\chicj$. \label{tab1} }
\begin{tabular}{lll}
\hline\hline
Decays & Helicity angles & Helicity amplitudes\\\hline
$\chicj(m)\to \Lambda^*(\lambda_R)\lmdb(\lambda_3)$ & $\theta_{[12]}^{[123]},\phi_{[12]}^{[123]}$ & $F^{\chicj}_{\lambda_R,\lambda_3}$\\
$\Lambda^*\to n(\lambda_1)K_S^0$ & $\theta_{[1]}^{[12]},\phi_{[1]}^{[12]}$ & $F^{\Lambda^*}_{\lambda_1,0}$\\\hline
$\chicj(m)\to\kst^+(\lambda_{+})K_S^0$ & $\theta_{[13]}^{[123]},\phi_{[13]}^{[123]}$ & $F^{\chicj}_{\lambda_{+},0}$\\
$K^{*+}\to\lmdb(\lambda_3)n(\lambda_1)$ & $\theta_{[3]}^{[13]},\phi_{[3]}^{[13]}$ & $F^{K^{*+}}_{\lambda_3,\lambda_1}$\\\hline
$\chicj(m)\to\bar N^{*-}(\lambda_{-})n(\lambda_1)$ & $\theta_{[23]}^{[123]},\phi_{[23]}^{[123]}$ & $F^{\chicj}_{\lambda_{-},\lambda_1}$\\
$\bar N^{*-}\to\lmdb(\lambda_3)K_S^0$ & $\theta_{[3]}^{[23]},\phi_{[3]}^{[23]}$ & $F^{\bar N^{*-}}_{\lambda_3,0}$\\\hline\hline
\end{tabular}
\end{center}
\end{table}
\subsection{Decay amplitude}

Decay amplitude for process (a) reads
\begin{equation}
A_1(m,\lambda_1,\lambda_3)=\sum_{\lambda_R} F^{\chicj}_{\lambda_R,\lambda_3}D^{J*}_{m,\lambda_R-\lambda_3}(\phi_{[12]}^{[123]},\theta_{[12]}^{[123]},0)
BW(m_{12})F^{\Lambda^*}_{\lambda_1,0}D^{J_R*}_{\lambda_R,\lambda_1}(\phi_{[1]}^{[12]},\theta_{[1]}^{[12]},0)
,
\end{equation}
where $D^J_{m,\lambda}(\phi,\theta,0)$ is Wigner-$D$ function with $J=0,1,2$ for $\chi_{cJ}$ states, $J_R$ is the spin of resonance $\Lambda^*$, and $BW$ denotes Breit-Wigner function.

Decay amplitude for the process (b) reads
\begin{eqnarray}
A_2(m,\lambda_1,\lambda_3)&=&\sum_{\lambda_+,\lambda_1',\lambda_3'} F^{\chicj}_{\lambda_+,0}D^{J*}_{m,\lambda_+}(\phi_{[13]}^{[123]},\theta_{[13]}^{[123]},0)
BW(m_{13})F^{K^{*+}}_{\lambda_3',\lambda_1'}D^{J_R*}_{\lambda_+,\lambda_3'-\lambda_1'}(\phi_{[3]}^{[13]},\theta_{[3]}^{[13]},0)\nonumber\\
&\times&D^{1\over2}_{\lambda_3',\lambda_3}(\phi'_2,\theta_2',\psi_2')D^{1\over2}_{\lambda_1',\lambda_1}(\phi'_3,\theta_3',\psi_3'),
\end{eqnarray}
where $J_R$ is the spin of $K^{*+}$. The angles ($\phi'_2,\theta_2',\psi_2'$) and ($\phi'_3,\theta_3',\psi_3'$) correspond to the rotation to align the $\bar\Lambda$ and $n$ helicity system to coincide with those defined in the process (a). These rotations carry the polarization of $\Lambda$ and $n$ into the helicity system of (a), so that ensures the coherent interference between the two decays. This issue has been addressed in the analyses \cite{lhcb,belle} and proved in Ref. \cite{chenh}.

Decay amplitude for process (c) reads
\begin{eqnarray}
A_3(m,\lambda_1,\lambda_3)&=&\sum_{\lambda_-,\lambda_1',\lambda_3'} F^{\chicj}_{\lambda_-,\lambda_1'}D^{J*}_{m,\lambda_--\lambda_1'}(\phi_{[23]}^{[123]},\theta_{[23]}^{[123]},0)
BW(m_{23})F^{\bar N^{*-}}_{\lambda_3',0}D^{J_R*}_{\lambda_-,\lambda_3'}(\phi_{[3]}^{[23]},\theta_{[3]}^{[23]},0)\nonumber\\
&\times&D^{1\over2}_{\lambda_3',\lambda_3}(\phi'_2,\theta_2',\psi_2')D^{1\over2}_{\lambda_1',\lambda_1}(\phi'_3,\theta_3',\psi_3'),
\end{eqnarray}
where the two factors $D^{1\over2}_{\lambda_3',\lambda_3}(\phi'_2,\theta_2',\psi_2')$ and $D^{1\over2}_{\lambda_1',\lambda_1}(\phi'_3,\theta_3',\psi_3')
$ correspond to the rotation of $\bar\Lambda$ and neutron helicities to coincide with those defined in the process $(a)$.

Helicity amplitude is expressed with partial waves in terms of $LS$-coupling scheme \cite{ref::chung, ref::chung2, ref::chung3}. For two-body decays with spin $J\to s+\sigma$, it follows
\begin{eqnarray}\label{chung_forma}
F^J_{\lambda,\nu} &=&\sum_{ls}\left ({2l+1\over 2J+1}\right )^{1/2}\langle l0S\delta|J\delta\rangle\langle s\lambda\sigma-\nu|S\delta\rangle g_{lS}r^l{B_l(r)\over B_l(r_0)},\nonumber \\
\end{eqnarray}
where $L$ and $S$ are the
orbital angular momentum and spin quantum numbers for the two-body
decay, $r$ is the magnitude of the relative momentum of the final
states, $r_0$ is the corresponding momentum at the $R$ nominal mass, $g_{ls}$ is a coupling constant
to be determined, and $B_L(r)$ is the Blatt-Weisskopf barrier factor
\cite{ref::blatt}.

\begin{figure}[htbp]
  \begin{center}
    \includegraphics[width=0.6\textwidth]{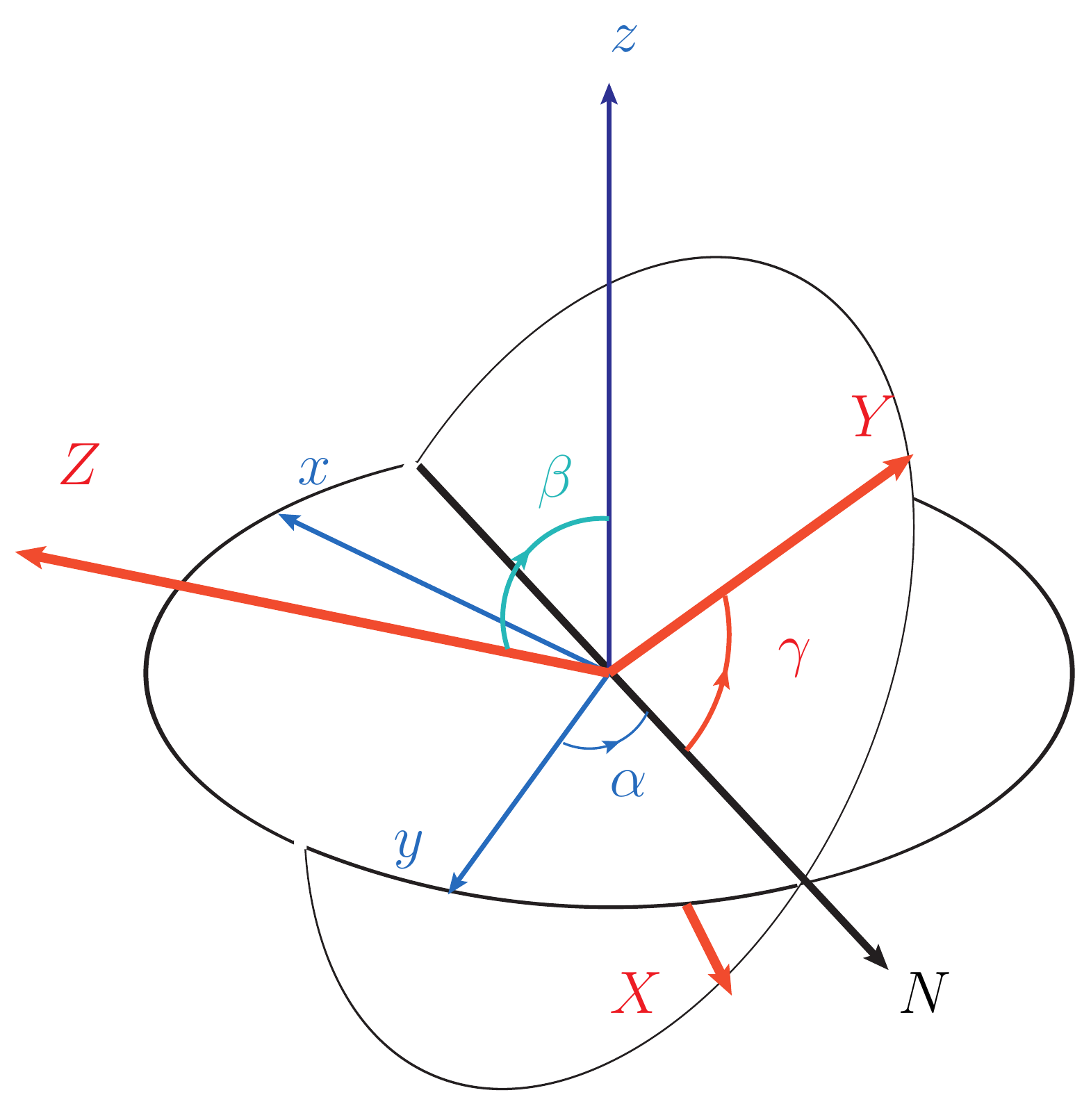}
     \caption{Illustration of rotations to carry the $n,K_S^0$ and $\Lambda$ orientations from the $\chi_{cJ}$ rest frame $xyz$ to the three body helicity system $XYZ$ by the three Euler angles $\alpha,\beta$ and $\gamma$.
    }
    \label{fig::eulerAngles}
  \end{center}
\end{figure}

The non-resonant decay is described with
the helicity amplitude constructed in the event system, in which the
$X$-$Y$ plane is chosen as the three-body decay plane, and the
$Z$-axis is defined as the normal to the decay plane. To describe the
$n,K_S^0$ and $\Lambda$ orientations in the $X$-$Y$-$Z$ system, one
needs to introduce three rotations to carry the momenta of final
states from the $\chi_{cJ}$ rest frame, $x$-$y$-$z$ to the three-body
helicity system, $X$-$Y$-$Z$ by three Euler angles,
$\alpha,~\beta$,and $\gamma$ as shown in Fig. \ref{fig::eulerAngles}.
The calculations of Euler angles can be found in \cite{ref::Devanathan}.

The amplitude is written as \cite{ref::helb3}, \begin{equation}
B_m=\sum_{\mu}D^{J*}_{m,\mu}(\alpha,\beta,\gamma)T_{\lambda_1,\lambda_2,\lambda_3} ~~,
\end{equation} where $m$ denotes the $z$-component of the $\chi_{cJ}$
spin in its rest frame. The decay amplitude $T_{\lambda_1,\lambda_2,\lambda_3}$ is a fit constant depending on the helicity values $\lambda_i$ of the
final state particles, and $\mu$ is the helicity of the $\chi_{cJ}$  in the helicity system fixed to the
three-body reference, which is summed over.
\subsection{Fit method}
To determine the decay coupling constants
$g_{ls}$ and decay amplitude $T$, we fit the model to the data events
by minimizing the function defined as 
\begin{equation} 
S=-(\ln\mathcal{L}_\text{data}- \ln \mathcal{L}_\text{bg}), 
\end{equation}
where the contributions from background events are subtracted from the
likelihood of observed data events $\mathcal{L}_\text{data}$, and the
likelihood function $\mathcal{L}$ for the $N$ data events is defined
with the amplitudes as 
\begin{equation} 
\mathcal{L}=\prod_{j=1}^N
\sum_{ m,\lambda_1,\lambda_3} {\rho^J_{m,m}\over \sigma_{\rm
MC}}\left|\sum_iA_i(m,\lambda_1,\lambda_3)+B_{m}\right|^2_j,
\end{equation}
where $\rho^J$ is the spin density matrix for the
$\chi_{cJ}$ production from the $\psi(3686)\to\gamma\chi_{cJ}$ decay,
and they are taken as $\rho^0=1,~\rho^1={1\over
4}\left(\begin{array}{clc} 1 &0 &0\\ 0 &2 &0\\ 0 &0 &1
\end{array}\right)$ and $\rho^2={3\over 20}\left(\begin{array}{clccc}
2 &0 &0 &0 &0\\ 0 &1 &0 & 0 &0\\0 &0 &2/3 &0 &0 \\0 &0 & 0 &1 &0\\ 0
&0 & 0 & 0 &2 \end{array}\right)$~\cite{chenh} for
$\chi_{c0},~\chi_{c1}$ and $\chi_{c2}$, respectively. The amplitude
summation $\sum_i A_i(...)$ is taken over all intermediate states, and
$[...]_j$ denotes the amplitude calculated with the angular
observables of $j$-th event. The normalization factor
$\sigma_\text{MC}$ is calculated as the average of amplitudes with a
sample of simulated MC phase space events.

To describe the data distributions with the model, we
use the 16 intermediate states [excited $\Lambda$:
  $\Lambda(1405),~\Lambda(1520),~\Lambda(1600),~\Lambda(1670)$,
  $\Lambda(1690)$, $\Lambda(1800)$, $\Lambda(1890)$, $\Lambda(2000)$, $\Lambda(2020)$, $\Lambda(2100)$,
  $\Lambda(2110)$, $\Lambda(2325)$,
  $\Lambda(2350)$; excited $N$: $N^*(2300)$; excited $K$:
  $K_2(2250)$], and the non-resonant decay. The masses and widths of
these states are fixed to the PDG values~\cite{ref::pdg2014} in the
fit. The possible interferences between the non-resonant process and
the processes with intermediate states are also included,
and parity conservation is required to reduce the
number of independent helicity amplitudes.

\bibliographystyle{JHEP}
\bibliography{references.bib}